\documentclass[a4paper,reqno]{amsart}
\pdfoutput=1

\usepackage{fancyhdr}
\usepackage{amsmath,amssymb,amsfonts,amsthm}
\usepackage{mathtools}
\usepackage{mathalfa}
\usepackage[T1]{fontenc}

\usepackage[dvipsnames]{xcolor}

\usepackage[hyphens]{url}
\usepackage{adjustbox}

\usepackage{hyperref}
\hypersetup{
    colorlinks=true,
    citecolor=Green,
    linkcolor=NavyBlue,
    filecolor=magenta,
    urlcolor=NavyBlue,
    }

\usepackage[capitalize]{cleveref}
\usepackage{upgreek}
\usepackage{bbm}
\usepackage{microtype}
\usepackage{tikz}
\usetikzlibrary{positioning,arrows}
\usepackage[most]{tcolorbox}
\usepackage{fontawesome5}
\usepackage{paralist}

\newcommand{\tikzxmark}{%
\tikz[scale=0.23,color=Mahogany] {
    \draw[line width=0.7,line cap=round] (0,0) to [bend left=6] (1,1);
    \draw[line width=0.7,line cap=round] (0.2,0.95) to [bend right=3] (0.8,0.05);
}}
\newcommand{\tikzcmark}{%
\tikz[scale=0.23, color=OliveGreen] {
    \draw[line width=0.7,line cap=round] (0.25,0) to [bend left=10] (1,1);
    \draw[line width=0.8,line cap=round] (0,0.35) to [bend right=1] (0.23,0);
}}

\usepackage{multirow}
\usepackage{caption} 
\usepackage{threeparttable}

\newcommand{\Bool}{\mathbf{2}}
\newcommand{\Unit}{\mathbf{1}}
\newcommand{\Empty}{\mathbf{0}}

\newcommand{\bff}{\mathsf{ff}}
\newcommand{\btt}{\mathsf{tt}}

\newcommand{\acc}{\mathsf{Acc}}

\newcommand{\Nat}{\mathbb{N}}
\newcommand{\isProp}{\mathsf{isProp}}
\newcommand{\Prop}{\mathsf{hProp}}
\newcommand{\isSet}{\mathsf{isSet}}
\newcommand{\Set}{\mathsf{hSet}}
\newcommand{\UU}{\mathcal{U}}

\newcommand{\defeq}{\vcentcolon\equiv}

\newcommand{\blank}{\_}
\newcommand{\dissum}{\mathbin{\uplus}}
\newcommand{\inl}{\mathsf{inl}}
\newcommand{\inr}{\mathsf{inr}}

\newcommand{\fstproj}{\mathsf{fst}}
\newcommand{\sndproj}{\mathsf{snd}}

\newcommand{\proptrunc}[1]{\|#1\|}
\newcommand{\truncP}[1]{|#1|}

\renewcommand{\iff}{\leftrightarrow}
\renewcommand{\implies}{\rightarrow}

\renewcommand{\O}{\ensuremath{\mathcal{O}}}

\newcommand{\cnf}{\mathsf{Cnf}}
\newcommand{\brouwer}{\mathsf{Brw}}
\newcommand{\bookord}{\mathsf{Ord}}

\newcommand{\bzero}{\mathsf{zero}}
\newcommand{\bsuc}{\mathsf{succ}}
\newcommand{\blimit}{\mathsf{limit}}
\newcommand{\bbisim}{\mathsf{bisim}}
\newcommand{\lzero}{\mathord\leq\mbox{-}\mathsf{zero}}
\newcommand{\ltrans}{\mathord\leq\mbox{-}\mathsf{trans}}
\newcommand{\lsuccmono}{\mathord\leq\mbox{-}\mathsf{succ}\mbox{-}\mathsf{mono}}
\newcommand{\lcocone}{\mathord\leq\mbox{-}\mathsf{cocone}}
\newcommand{\llimiting}{\mathord\leq\mbox{-}\mathsf{limiting}}

\newcommand{\toincr}[1]{\xrightarrow{#1}}

\newcommand{\isZero}{\mathsf{isZero}}
\newcommand{\Code}{\mathsf{Code}}
\newcommand{\toCode}{\mathsf{toCode}}
\newcommand{\fromCode}{\mathsf{fromCode}}

\newcommand{\osuc}[1]{#1 \dissum \Unit}
\newcommand{\osup}[1]{\lim #1}

\newcommand{\simplesup}{\mathsf{sup}}

\newcommand{\fst}{\mathsf{left}}

\newcommand{\TT}{\mathcal T}

\newcommand{\fatplus}{\raisebox{0ex}{\tikz\filldraw[black,x=.7pt,y=.7pt] (0,0) -- ++(3,0) -- ++(0,3) -- ++(1,0) -- ++(0,-3) -- ++(3,0) -- ++(0,-1) -- ++(-3,0) -- ++(0,-3) -- ++(-1,0) -- ++(0,3) -- ++(-3,0) -- cycle;}}

\newcommand{\tz}{\mathrm 0 }
\newcommand{\tom}[2]{{\upomega}^{#1} \, \fatplus \, {#2}}

\newcommand{\tone}{1}

\newcommand{\N}{\mathbb N}

\newtcolorbox{summary}{%
	colback=white,
	borderline={1pt}{-2pt}{black},
	left=4pt,
	right=4pt,
	top=4pt,
	bottom=4pt,
	title={\textit{\textbf{Summary of results}}},
	boxrule=.8pt,
}

\newcommand{\summarySpace}{\vspace*{.0cm}}

\newcommand{\iszero}{\mathsf{is}\mbox{-}\mathsf{zero}}

\newcommand{\isstrongsuc}{\mathsf{is}\mbox{-}\mathsf{str}\mbox{-}\mathsf{suc}}
\newcommand{\issucof}[2]{#1 \; \mathsf{is}\mbox{-}\mathsf{suc}\mbox{-}\mathsf{of} \; #2}
\let\isupsucof\issucof
\newcommand{\isstrongsucof}[2]{#1 \; \mathsf{is}\mbox{-}\mathsf{str}\mbox{-}\mathsf{suc}\mbox{-}\mathsf{of} \; #2}

\newcommand{\issupof}[2]{#1 \; \mathsf{is}\mbox{-}\mathsf{sup}\mbox{-}\mathsf{of} \; #2}
\let\isupsupof\issupof

\newcommand{\islim}{\mathsf{is}\mbox{-}\N\mbox{-}\mathsf{lim}}
\newcommand{\islimof}[2]{#1 \; \mathsf{is}\mbox{-}\N\mbox{-}\mathsf{lim}\mbox{-}\mathsf{of} \; #2}
\newcommand{\isgenerallim}{\mathsf{is}\mbox{-}\mathsf{general}\mbox{-}\mathsf{lim}}

\newcommand{\CtoB}{\mathsf{CtoB}}
\newcommand{\BtoO}{\mathsf{BtoO}}

\newcommand{\formalisedPartial}{\faCog}
\newcommand{\formalisedFull}{\faCogs}
\newcommand{\flinkPart}[1]{\href{#1}{\formalisedPartial}}
\newcommand{\flinkFull}[1]{\href{#1}{\formalisedFull}}

\newcommand{\LEM}{\ensuremath{\mathsf{LEM}}}
\newcommand{\MP}{\ensuremath{\mathsf{MP}}}
\newcommand{\LPO}{\ensuremath{\mathsf{LPO}}}
\newcommand{\WLPO}{\ensuremath{\mathsf{WLPO}}}

\newcommand{\Dec}{\mathsf{Dec}}
\newcommand{\Stable}{\mathsf{Stable}}
\newcommand{\Splits}{\mathsf{Splits}}

\newcommand{\jump}[1]{{#1}^{\uparrow}}
\newcommand{\unjump}[2]{{#1}^{\downarrow #2}}

\newtheorem{theorem}{Theorem}
\newtheorem{lemma}[theorem]{Lemma}
\newtheorem{corollary}[theorem]{Corollary}
\theoremstyle{definition}
\newtheorem{definition}[theorem]{Definition}

\theoremstyle{remark}
\newtheorem{remark}[theorem]{Remark}

\newcommand{\useBoxCref}{	
	\crefname{theorem}{Thm}{Thms}
	\crefname{lemma}{Lem}{Lems}
	\crefname{corollary}{Cor}{Cors}
}
\author{Nicolai Kraus \and Fredrik Nordvall Forsberg \and Chuangjie Xu}
\title{Type-Theoretic Approaches to Ordinals}

\begin{document}
\begin{abstract}
	In a constructive setting, no concrete formulation of ordinal numbers can simultaneously have all the properties one might be interested in; for example, being able to calculate limits of sequences is constructively incompatible with deciding extensional equality.
	Using homotopy type theory as the foundational setting, we develop an abstract framework for ordinal theory and establish a collection of desirable properties and constructions.
	We then study and compare three concrete implementations of ordinals in homotopy type theory: first, a notation system based on Cantor normal forms (binary trees); second, a refined version of Brouwer trees (infinitely-branching trees); and third, extensional well-founded orders.
	
	Each of our three formulations has the central properties expected of ordinals, such as being equipped with an extensional and well-founded ordering as well as allowing basic arithmetic operations, but they differ with respect to what they make possible in addition.
	For example, for finite collections of ordinals, Cantor normal forms have decidable properties, but suprema of infinite collections cannot be computed. In contrast, extensional well-founded orders work well with infinite collections, but the price to pay is that almost all properties are undecidable.
	Brouwer trees, implemented as a quotient inductive-inductive type to ensure well-foundedness and extensionality, take the sweet spot in the middle by combining a restricted form of decidability with the ability to work with infinite increasing sequences.
	
	Our three approaches are connected by canonical order-preserving functions from the ``more decidable'' to the ``less decidable'' notions, i.e.\ from Cantor normal forms to Brouwer trees, and from there to extensional well-founded orders.
	We have formalised the results on Cantor normal forms and Brouwer trees in cubical Agda, while extensional well-founded orders have been studied and formalised thoroughly by Escard\'{o} and his collaborators.
	Finally, we compare the computational efficiency of our implementations with the results reported by Berger.

	\vspace*{1em}

\end{abstract}

\maketitle

\newpage

\setcounter{tocdepth}{2}

\tableofcontents
\newpage

\section{Introduction}

Ordinal numbers are an important tool in modern mathematics and
proof theory, employed for example for showing termination of
processes~\cite{dershowitz:termination,Floyd:1967},
the semantics of inductive definitions~\cite{aczelinductive,Dybjer99afinite}, and
justifying recursion, as used in many papers by Berger and others on
realisability and program extraction in the presence of induction and
coinduction~\cite{realInd,bergerPPP,IFP}.
In these applications of ordinals, the metatheory is typically based on classical logic. The applications, however, are of interest also in constructive mathematics, and so, it would be of great benefit to develop constructive approaches to ordinals which are strong enough to handle such applications. In this article, we use type theory to develop such constructive approaches, and show that they also cover other important aspects of ordinals, such as their arithmetic theory, generalising the one of the natural numbers, and the existence of suprema of potentially unbounded sequences of ordinals.

\subsection{Three Approaches to Ordinal Numbers} \label{subsec:three-approaches}

In classical set theory, there are various equivalent representations of ordinals, and, whenever one representation is more convenient than another, one can freely switch between them.
However, switching to a different representation often requires the law of excluded middle or other constructively unavailable principles.
Therefore, it is unsurprising that the situation is more challenging in a constructive setting.
Different representations of ordinals are no longer equivalent, and it is easy to see that there cannot be a single formulation of ordinals that makes it possible to prove all the properties and perform all the constructions that the various applications require.
For example, consider a binary sequence $s$, i.e.\ a function from the natural numbers into the set $\{0, 1\}$; if we had a formulation of ordinals that allowed us to calculate the limit $x$ of the sequence $s$ \emph{and} decide extensional equality of $x$ with $0$, then this would amount to checking whether the sequence $s$ is constantly $0$. This, however, is exactly Bishop's \emph{weak limited principle of omniscience}~\cite{bishop1967}, an axiom that is generally not assumed in constructive mathematics.

When using or developing ordinals in a constructive setting, one is for this reason forced to make compromises and give up some desirable properties, and the choice will naturally depend on the anticipated applications.
This explains why several different constructions of ordinals have been studied in the literature. It is also not unusual to see tailor-made inductive definitions replace applications of ordinals and transfinite induction, with the consequence that basic results are established over and over again, for each inductive definition. Instead, in this article, we will consider the following approaches to ordinals:
\begin{itemize}
	\item One approach to develop ordinal theory is to use ``syntactic'' ordinal notation
	systems~\cite{buchholz:notation,schuette:book,takeuti:book}. 
	Such systems are popular with proof theorists,
	as their concrete character typically
	means that equality and the order relation on ordinals are
	decidable. However, truly infinitary operations such as calculating
	limits of infinite families of notations are usually not constructible.
	\item Another approach to ordinals, popular in the functional programming community and based on notation systems by Church~\cite{church:1938} and Kleene~\cite{kleene:notation-systems}, 
	is to consider \emph{Brouwer trees} $\O$ inductively generated by zero,
	successor, and a supremum constructor
	\begin{equation} \label{eq:fake-sup}
	\mathsf{sup} : (\N \to \O) \to \O
	\end{equation}
	which forms a new tree for every countable sequence of
	trees~\cite{brouwer:trees,coquand:ord-in-tt,hancock:thesis}.
	By the inductive nature of the definition, constructions on trees can be
	carried out by giving one case for zero, one for successors, and one
	for suprema, just as in the classical theorem of transfinite
	induction.
	Of course, when allowing infinite sequences, extensional equality cannot be checked algorithmically.
	\item Yet another approach to ordinals is to consider collections of extensional well-founded orders satisfying transitivity,
	representing a variation on the classical set-theoretical axioms that is more
	suitable for a constructive treatment~\cite{taylor:ordinals}.
	When pursuing a development of ordinals based on such orders without further conditions, one naturally gives up all non-trivial notions of decidability -- it even becomes impossible to check whether a given order is zero, a successor, or a limit.
	Nevertheless, many operations can still be defined on the collection of all such orders,
	and properties such as well-foundedness can still be proven. This is
	also the notion of ordinal most closely related to the traditional
	notion, and thus, in a classical setting, the formulation which most obviously corresponds to the established literature.
\end{itemize}


Is it possible to specify what a proper definition of ordinals in a constructive setting is, how the different approaches fulfil the specification, which properties they lack, and how they are connected to each other?
Can constructions and ideas
be transported from one setting to another --- e.g., do the arithmetic
operations constructed for one notion of ordinals obey the same rules as the
arithmetic operations defined for another notion?
In order to develop one possible precise answer to these questions, we work in \emph{homotopy type theory}~\cite{hott-book}
and suggest an abstract framework of ordinals in which the various desirable properties and operations can be formulated.
We also study three concrete constructions, representing the three approaches
above, which become instances of our general abstract framework.

The representative of the class of ``syntactic'' ordinal notation systems that we develop is based on Cantor
normal forms using unlabelled binary trees.
Our concrete definition ensures that we have no ``junk'' terms, i.e.\ each element of our type $\cnf$ of Cantor normal forms denotes an actual ordinal.
There are several reasonable ways in which such a type can be defined that can be shown to be equivalent to each other~\cite{NFXG:three:ord}; the construction that we choose defines $\cnf$ as a subset of the type of binary trees.

When defining Brouwer trees as an inductive type with constructors for zero, successor, and a third constructor $\mathsf{sup}$ as in \eqref{eq:fake-sup},
it is a priori wishful thinking to call the last constructor a ``supremum'';
$\mathsf{sup}(s)$ does not faithfully represent the suprema
of the sequence $s$, since we do not have that e.g.\
$\mathsf{sup}(s_0, s_1, s_2, \ldots) = \mathsf{sup}(s_1, s_0, s_2, \ldots)$
--- each sequence gives rise to a new tree, rather than
identifying trees that would be supposed to represent the same supremum.
Fortunately, this shortcoming can be fixed in homotopy type theory via a \emph{higher} or \emph{quotient inductive-inductive type} $\brouwer$,
combining induction-induction with the idea of higher inductive
types~\cite{cubicalhits,lumsdaine:hits}.
While we naturally cannot derive decidable equality for the type $\brouwer$, 
we retain the possibility of classifying an ordinal as
a zero, a successor or a limit.

Finally, the idea of extensional well-founded orders was transferred to the setting of homotopy type theory in the ``HoTT
book''~\cite[Chp~10]{hott-book}, and significantly extended by Escard\'{o} and his collaborators~\cite{escardo:agda-ordinals}.
The approach is to define $\bookord$ to be the type of pairs $(X,<)$, where the latter is a propositionally-valued, transitive, extensional, and well-founded relation.
While $\bookord$ lacks all forms of non-trivial decidability, it is better suited for constructions involving infinite families of ordinals than $\cnf$ or $\brouwer$.

\newcommand{\LPOref}{\hyperlink{ref:tabLPO}{\tnote{*}}}
\begin{table}[t]
\begin{adjustbox}{center}
\begin{threeparttable}[t]
\begin{tabular}{| l || c c c|| c |} 
 \hline
  Property & $\cnf$ & $\brouwer$ & $\bookord$ & More details\\ 
 \hline\hline
 $x < y \leq z \to x < z$ & \tikzcmark & \tikzcmark & \tikzcmark & \multirow{2}{*}{beginning of \cref{sec:axiomaticapproach}} \\
 $x \leq y < z \to x < z$ & \tikzcmark & \tikzcmark & \tikzxmark & \\
 \hline
 order is well-founded & \tikzcmark & \tikzcmark & \tikzcmark & \multirow{2}{*}{\cref{subsec:ext-wf}}\\ 
 order is extensional & \tikzcmark & \tikzcmark & \tikzcmark & \\ 
 \hline
 has zero and successors & \tikzcmark & \tikzcmark & \tikzcmark & \multirow{5}{*}{\cref{subsec:class}}\\
 has finite suprema (binary joins) & \tikzcmark & \tikzxmark & \tikzcmark & \\
 has limits of strictly increasing $\N$-sequences & \tikzxmark & \tikzcmark & \tikzcmark & \\
 has suprema of small families & \tikzxmark & \tikzxmark & \tikzcmark & \\
 can classify as zero, successor, or limit & \tikzcmark & \tikzcmark & \tikzxmark & \\ 
 \hline
 has addition & \tikzcmark & \tikzcmark & \tikzcmark & \multirow{5}{*}{\cref{subsec:abstract-arithmetic}}\\
 has multiplication & \tikzcmark & \tikzcmark & \tikzcmark & \\
 has (partial) exponentiation  & \tikzcmark & \tikzcmark &  & \\
 has subtraction & \tikzcmark & \tikzxmark\LPOref & \tikzxmark & \\
 has division & \tikzcmark & \tikzxmark & \tikzxmark & \\
 \hline
 $x < \omega$ is decidable & \tikzcmark & \tikzcmark & \tikzxmark & \multirow{4}{*}{\cref{subsec:Decidability}} \\ 
 $x < y$ is decidable & \tikzcmark & \tikzxmark\LPOref & \tikzxmark & \\
 $x\leq y$ splits as $(x<y) \uplus (x=y)$ & \tikzcmark & \tikzxmark\LPOref  & \tikzxmark & \\
 is trichotomous & \tikzcmark & \tikzxmark\LPOref  & \tikzxmark & \\
 \hline
 computational efficiency & (good) & (medium) & (none) & \cref{sec:bench}\\
 \hline
\end{tabular}
\begin{tablenotes}
	\item[\hypertarget{ref:tabLPO}{*}] Each of these properties is equivalent to Bishop's \emph{limited principle of omniscience} (\LPO), cf.~\cref{subsec:taboos}.
\end{tablenotes}
\end{threeparttable}
\end{adjustbox}
\caption{Summary of how the three notions of ordinals compare.}
\label{table:1}
\end{table}

All in all, each of the approaches above gives quite a different feel to
the ordinals they represent: Cantor normal forms emphasise syntactic
manipulations, Brouwer trees how every ordinal can be classified as a
zero, successor or limit, and extensional well-founded orders the set-theoretic properties of ordinals.
It turns out that there are canonical embeddings of the ``more'' into the ``less decidable'' notions,
i.e.\ we have functions from Cantor normal forms to Brouwer trees and from Brouwer trees to extensional well-founded orders.
We study whether, or under which conditions, these functions preserve arithmetic operations, commute with limits, and are simulations.
Inspired by Berger's results on the computational efficiency of implementations of ordinals~\cite{Berger01a}, we also show how $\cnf$ and $\brouwer$ perform when implemented in Cubical Agda~\cite{VMA:cubical:agda}.
Note that we cannot compute in a meaningful way with $\bookord$.
A summary of how the three notions of ordinals compare can be found in \cref{table:1}.

%

\subsection{Related Work}

The development of ordinals in constructive mathematics has a rich history~\cite{brouwer:trees, church:1938, gentzen-zahlentheorie,  heyting-infinitistic, kleene:notation-systems, Martin-Lof1970-MARNOC, Mines/R/R:1988}. As mentioned above, one well-known constructive notion of ordinal is given by extensional well-founded relations. However, the transitivity
\begin{equation}
\label{eq:trans}
x \leq y < z \implies x<z
\end{equation}
fails because it implies excluded middle. Taylor~\cite{taylor:ordinals,taylor-book} recovers it by introducing plumpness, which essentially restricts to the subclass for which the property~(\ref{eq:trans}) holds hereditarily. We do not explicitly study plump ordinals in this paper, but the transitivity~(\ref{eq:trans}) holds for both Cantor normal forms and Brouwer trees. In their recent work~\cite{CLN-constructive-ordinals}, Coquand, Lombardi and Neuwirth develop another constructive theory of ordinals. They start with a structure of certain linear orders, called $\mathfrak{F}$-orders, to describe the desirable properties of ordinals including the transitivity~(\ref{eq:trans}). Then they inductively construct a set $\mathbf{Ord}$ of ordinals and prove that $\mathbf{Ord}$ is initial in the category of $\mathfrak{F}$-orders. In this way, they show that their constructive ordinals satisfy the desirable properties constructively. Our abstract axiomatic framework introduced in~\cref{sec:axiomaticapproach} is similar to their structure of $\mathfrak{F}$-orders. But it is more general (e.g., no assumption like~(\ref{eq:trans})) because it is for relating and comparing our three different approaches to ordinals.

Several formalisations of ordinals and ordinal notation systems exist in the literature. Escard\'o and his collaborators develop many results of extensional well-founded relations in homotopy type theory, and have formalised  them in Agda~\cite{escardo:agda-ordinals}. The theory of ordinals below $\varepsilon_0$ based on various representations has been developed in some formal systems. For the representation of Cantor normal forms with coefficients, Manolios and Vroon~\cite{MV:ord:acl2} work in ACL2, Cast\'eran and Contejean \cite{CC:ord:coq} and Grimm~\cite{grimm:ord:coq} in Coq, and Shinkarov~\cite{shinkarov-agda} in Agda. Blanchette, Fleury and Traytel~\cite{BFT:nested:mset} work with the representation of hereditary multisets in Isabelle/HOL. One of our approaches to ordinals is based on Cantor normal forms without coefficients. In this paper, we additionally prove that the arithmetic operations on Cantor normal forms are uniquely correct with respect to our abstract axiomatisation.

In the work on ordinals below $\varepsilon_0$~\cite{CC:ord:coq,grimm:ord:coq,NFXG:three:ord, shinkarov-agda}, one derives/implements transfinite induction directly from the selected representation of ordinals. Berger~\cite{Berger01a} instead extracts a program from Gentzen's proof~\cite{gentzen-transfinite} of transfinite induction up to $\varepsilon_0$. Gentzen's proof involves nesting of implications of bounded depth. Therefore, the extracted program contains functionals of arbitrarily high types (with respect to the finite type structure). Using the extracted program, one obtains higher type primitive recursive definitions of the fast growing hierarchy and tree ordinals of height below $\varepsilon_0$. Berger compares the transfinite recursive implementation of the Hardy hierarchy and the one via the extracted higher type program, and observes that the latter is much faster. Inspired by Berger's experiment, we compare the computational efficiency of our notions of ordinals in~\cref{sec:bench}.

\subsection{Agda Formalisation}

As indicated above, we have formalised the material on Cantor normal forms and Brouwer
trees in cubical Agda~\cite{VMA:cubical:agda}:
\begin{itemize}
\item Git repository of the Agda code: \href{https://bitbucket.org/nicolaikraus/constructive-ordinals-in-hott/}{\nolinkurl{bitbucket.org/nicolaikraus/constructive-ordinals-in-hott/}}
\item DOI of the archived Agda code: \href{https://doi.org/10.5281/zenodo.7657456}{10.5281/zenodo.7657456}
\item HTML rendering: \href{https://cj-xu.github.io/agda/type-theoretic-approaches-to-ordinals/}{\nolinkurl{cj-xu.github.io/agda/type-theoretic-approaches-to-ordinals/}}
\end{itemize}
While the development in the Git repository listed above is ongoing and not restricted to the current paper, the HTML rendering and the archived version are snapshots of the repository at the time of writing. 
To complement the formalisation, we also refer to Escard\'{o}'s Agda library~\cite{escardo:agda-ordinals} that consists of many
results on extensional well-founded orders in HoTT.
In the paper, we have marked theorems whose proofs we have formalised, or partly formalised, using the
symbols \formalisedFull{} and \formalisedPartial{}
respectively; they are also clickable links to the corresponding
machine-checked statement in the HTML rendering of our Agda code.

Our formalisation builds on the agda/cubical library~\cite{agdacubical} and type checks using Agda version 2.6.3.
It uses the \texttt{\{-\# TERMINATING \#-\}} pragma to
work around a limitation of the termination checker of Agda: recursive calls
hidden under a propositional truncation are not seen to be
structurally smaller. 
While we believe that such recursive calls when proving a proposition are justified by Dijkstra's eliminator presentation~\cite{gabe:phd}, it would be non-trivial to reduce our mutual definitions to eliminators.

\subsection{History of this Paper}

An earlier, short version of this paper appeared in the proceedings of the conference
\emph{Mathematical Foundations of Computer Science}~\cite{MFCS}.
New in this version is a thorough discussion of the constructive aspects and decidability results (\cref{subsec:Decidability}) of the various approaches to ordinals via \emph{constructive taboos} (cf.~\cref{subsec:taboos}).
In particular, we rigorously study the decidable and undecidable properties of Brouwer trees (\cref{subsec:Decidability}) and connect the preservation of limits of the embedding $\cnf \to \brouwer$ with Markov's principle (cf.~\cref{lem:f-arith,thm:CtoB-preserves-fundamental-seqs}).
In addition to providing a complete Agda formalisation (except for extensional well-founded orders), we benchmark the efficiency of computing with Cantor normal forms and Brouwer trees
(\cref{sec:bench}), inspired by Berger's benchmarking of
ordinal recursive versus higher type programs~\cite{Berger01a}.

\section{Preliminaries}

In this section, we introduce concepts and notation that we are going to use
in the rest of the paper.

\subsection{Concepts of Homotopy Type Theory}

We work in and assume basic familiarity with homotopy type theory
(HoTT), i.e.\ Martin-L\"of type theory extended with higher inductive
types and the univalence axiom~\cite{hott-book}. The central concept
of HoTT is the Martin-L\"of identity type, which we write as $a = b$
--- we write $a \equiv b$ for definitional equality. We use Agda
notation $(x : A) \to B(x)$ for the type of dependent functions, and
write simply $A \to B$ if $B$ does not depend on $x : A$.
We further write $A \leftrightarrow B$ for ``if and only if'', i.e.\
for functions in both directions $A \to B$ and $B \to A$.
If the type in the domain can be inferred from context, we may simply
write $\forall x. B(x)$ for $(x : A) \to B(x)$.  Freely occurring
variables are assumed to be $\forall$-quantified.

We denote the type of dependent pairs by $\Sigma(x : A).B(x)$, and its
projections by $\fstproj$ and $\sndproj$. We write $A\times B$ if $B$
does not depend on $x:A$.  We write $\UU$ for a universe of types; we
assume that we have a cumulative hierarchy $\UU_i : \UU_{i+1}$ of such
universes closed under all type formers, but we will leave universe
levels typically ambiguous.

We call a type $A$ a \emph{proposition}, $\isProp(A)$, if all elements of $A$ are
equal, i.e.\ if $(x : A) \to (y : A) \to x = y$ is provable. We write
$\Prop = \Sigma(A : \UU).\isProp(A)$ for the type of
propositions, and we implicitly insert a first projection if
necessary, e.g.\ for $A : \Prop$, we may write $x : A$ rather than
$x : \fstproj(A)$.  A type $A$ is a \emph{set}, $\isSet(A)$, if
$\isProp(x = y)$ for every $x, y : A$. We write  $\Set = \Sigma(A : \UU).\isSet(A)$ for the type of sets, again with the first projection implicit when necessary.

We denote \emph{propositional truncation} of a type $A$ by $\proptrunc{A}$, which is the smallest proposition with a function from $A$. In particular, by $\exists(x : A).B(x)$, we mean the propositional truncation
of $\Sigma(x : A).B(x)$, i.e., we have $\exists(x : A).B(x) : \Prop$, and if
$(a, b) : \Sigma(x : A).B(x)$ then $|(a, b)| : \exists(x : A).B(x)$.
The elimination rule of $\exists(x : A).B(x)$ only allows to define
functions into propositions. By convention, we write $\exists k. P(k)$
for $\exists(k : \N).P(k)$.  Finally, we write $A \dissum B$ for the
sum type, $\Empty$ for the empty type, $\Unit$ for the type with
exactly one element $\ast$, and $\Bool$ for the type with two elements
$\bff$ and $\btt$.

\subsection{Constructive Taboos}
\label{subsec:taboos}

When we are unable to perform a certain construction or prove a theorem formulated as a type $A$,
we want to understand why it is seemingly impossible to define an element of $A$.
An obvious approach is attempting to derive a contradiction from the assumption that $A$ holds, i.e.\ prove $A \to \Empty$, a type that we also denote by $\neg A$ (``not $A$'').
However, this will often not be possible either in a constructive setting, since many interesting statements $A$ are consistent and may actually turn out to be true in a classical theory.
Therefore, we may be forced to replace the empty type by a less ambitious goal $B$; something that is known from models to not be provable in the type theory we work in, or something that is simply generally undesirable in a constructive setting.
We call such a type $B$ a \emph{constructive taboo}; it is also sometimes known as a Brouwerian counterexample~\cite{varieties}.

Obviously, the empty type $\Empty$ is a taboo in all interesting (non-trivial) settings, so it is technically also a constructive taboo.
Another very prominent constructive taboo is the \emph{law of excluded middle} $\LEM$, stating that every proposition is either true or false:
\begin{equation}
\LEM \defeq \forall (P : \Prop). P \dissum \neg P.
\end{equation}

Several other taboos that we consider talk about binary sequences.
The first is Bishop's \emph{limited principle of omniscience} $\LPO$ \cite{bishop1967}, stating that every binary sequence is either constantly $\bff$ or somewhere $\btt$:
\begin{equation}
\LPO \defeq \forall (s : \Nat \to \Bool). (\forall n. s_n = \bff) \dissum (\exists n. s_n = \btt).
\end{equation}
The weakened version, known as the \emph{weak limited principle of omniscience} $\WLPO$, states that it is decidable whether a sequence is constantly $\bff$:
\begin{equation}
\WLPO \defeq \forall (s : \Nat \to \Bool). (\forall n. s_n = \bff) \dissum \neg(\forall n. s_n = \bff).
\end{equation}

Finally, \emph{Markov's principle} $\MP$ says that, if a sequence is not constantly $\bff$, then it is $\btt$ somewhere:
\begin{equation}
\MP \defeq  \forall (s : \Nat \to \Bool). \neg(\forall n. s_n = \bff) \to (\exists n. s_n = \btt).
\end{equation}
We always have
\(
(\forall n. s_n = \bff) \; \leftrightarrow \; \neg(\exists n. s_n = \btt).
\)
If we view a binary sequence $s : \N \to \Bool$ as representing a semidecidable property (cf.~\cite{bauer2006first}),
then $\LPO$ says that every semidecidable property is decidable ($P \dissum \neg P$), while $\WLPO$ says that every semidecidable property is weakly decidable ($\neg P \dissum \neg\neg P$), and $\MP$ postulates that every semidecidable property is stable ($\neg\neg P \to P$).
It is thus not surprising, and well known, that we have:
\begin{lemma}[\flinkFull{https://cj-xu.github.io/agda/type-theoretic-approaches-to-ordinals/index.html\#Lemma-1}]
  \label{lem:lpo-equiv-lpo-mp}
	The limited principle of omniscience is as strong as the weak limited principle of omniscience and Markov's principle combined:
	\begin{equation}
	\LPO \; \leftrightarrow \; \WLPO \times \MP. 
	\end{equation}
	\qed
\end{lemma}
Note that the distinction between $\exists$ and $\Sigma$ is inessential in the formulation of the above taboos.
This is shown by the following lemma (see e.g.~Escard\'o~\cite{escardo:IAS:lpo} and Escard\'o and Xu~\cite[\S{}3.1]{escardoXu_continuityInconsistent}):
\begin{lemma}[\flinkFull{https://cj-xu.github.io/agda/type-theoretic-approaches-to-ordinals/index.html\#Lemma-2}]
  \label{lemma:sigma-lpo-vs-lpo}
	For any sequence $s : \N \to \Bool$, we have
	\begin{equation}
	(\exists n. s_n = \btt) \to \Sigma(n : \N). s_n = \btt.
	\end{equation}
	In particular, if we assume $\LPO$, then a given sequence is either constantly $\bff$ or we concretely get an $n : \N$ where it is not.
\end{lemma}
\begin{proof}
	Refining the type of $n$ such that $s_n = \btt$ to the type of \emph{minimal} $n$ with this property, we get a proposition and can eliminate out of the truncation.
	In detail, we construct a function
	\begin{equation}
	\forall n. s_n = \btt \to \left(\Sigma(n : \N). (s_n = \btt) \times \Pi(k < n). s_k = \bff\right)
	\end{equation}
	that searches the minimal index where a sequence is positive.
	Using that the target is a proposition, we precompose this function with the eliminator of the truncation.
	Finally, we compose with the projection function forgetting that $n$ is minimal.
\end{proof}

\section{Three Constructions of Types of Ordinals}
\label{sec:three-constructions-overview}

We consider three concrete notions of ordinals in this paper, together
with their order relations $<$ and $\leq$.
The first notion is the one of \emph{Cantor normal forms}, written
$\cnf$, whose order is decidable.  The second, written $\brouwer$, are
\emph{Brouwer Trees}, implemented as a higher inductive-inductive
type.  Finally, we consider the type $\bookord$ of ordinals that were
studied in the HoTT book~\cite{hott-book}, whose order is undecidable,
in general.
In the current section, we briefly give the three definitions and
leave the discussion of results for afterwards.

\subsection{Cantor Normal Forms as a Subset of Binary Trees}
\label{subsec:CNF}

In classical set theory, every ordinal $\alpha$ can be written
uniquely in Cantor normal form
\begin{equation}
  \label{eq:cnf-set-theory}
\alpha = \omega^{\beta_1} + \omega^{\beta_2} + \cdots + \omega^{\beta_n}
\;
\text{with }
\beta_1 \geq \beta_2 \geq \cdots \geq \beta_n
\end{equation}
for some natural number $n$ and ordinals $\beta_i$. If
$\alpha < \varepsilon_0$, then $\beta_i < \alpha$, and we can
represent $\alpha$ as a finite binary tree (with a condition) as
follows~\cite{buchholz:notation,CC:ord:coq,grimm:ord:coq,NFXG:three:ord}. Let
$\TT$ be the type of unlabeled binary trees, i.e.\ the inductive type
with suggestively named constructors $\tz : \TT$ and
$\tom{-}{-} : \TT \times \TT \to \TT$.  Let the relation $<$ be the
\emph{hereditary lexicographical order}, i.e.\ generated by the following
clauses:
 \begin{align}
   & \tz < \tom a b \\
   & a < c \to \tom a b < \tom c d \\
   & b < d \to \tom a b < \tom a d.
  \end{align}
We have the map $\fst:\TT \to \TT$ defined by $\fst(\tz) \defeq \tz$
and $\fst(\tom{a}{b}) \defeq a$ which gives us the left subtree (if it
exists) of a tree.  A tree is a \emph{Cantor normal form} (CNF) if,
for every $\tom{s}{t}$ that the tree contains, we have
$\fst(t) \leq s$, where $s \leq t \defeq (s < t) \dissum (s = t)$;
this enforces the condition in \eqref{eq:cnf-set-theory}. For
instance, both trees $\tone \defeq \tom{\tz}{\tz}$ and
$\omega \defeq \tom{\tone}{\tz}$ are CNFs.  Formally, the predicate
$\mathsf{isCNF}$ is defined inductively by the two clauses
 \begin{align}
  & \mathsf{isCNF}(\tz) \\
  & \mathsf{isCNF}(s) \to \mathsf{isCNF}(t) \to \fst(t) \leq s \to \mathsf{isCNF}(\tom{s}{t}).
 \end{align}
We write $\cnf \defeq \Sigma(t : \TT). \mathsf{isCNF}(t)$ for the type
of Cantor normal forms.  We often omit the proof of
$\mathsf{isCNF}(t)$ and call the tree $t$ a CNF if no confusion is
caused.

\subsection{Brouwer Trees as a Quotient Inductive-Inductive Type}
\label{subsec:Brouwer-as-HIIT}

As discussed in the introduction, \emph{Brouwer trees} (or \emph{Brouwer ordinal trees}) in functional programming are often
inductively generated by the usual constructors of natural numbers
(zero and successor) and a constructor that gives a
Brouwer tree for every sequence of Brouwer trees.  To state a refined
(\emph{correct} in a sense that we will make precise and prove)
version, we need the following notions:

Let $A$ be a type and $\prec \, : A \to A \to \Prop$ be a binary
relation.  If $f$ and $g$ are two sequences $\N \to A$, we say that
$f$ is \emph{simulated by} $g$, written $f \precsim g$, if
$f \precsim g \defeq \forall k. \exists n. f(k) \prec g(n)$.  We say
that $f$ and $g$ are \emph{bisimilar} with respect to $\prec$, written
\mbox{$f \approx^{\prec} g$}, if we have both $f \precsim g$ and
$g \precsim f$.  A sequence $f : \N \to A$ is \emph{increasing} with
respect to $\prec$
if we have \mbox{$\forall k. f(k) \prec f(k+1)$}.  We write
$\N \toincr{\prec} A$ for the type of $\prec$-increasing
sequences. Thus an increasing sequence $f$ is a pair
$f \equiv (\overline f, p)$ with $p$ witnessing that $\overline f$ is
increasing, but we keep the first projection implicit and write $f(k)$
instead of $\overline f(k)$.

Our type of Brouwer trees is a \emph{quotient inductive-inductive
  type} \cite{Altenkirch2018}, where we simultaneously construct the
type $\brouwer : \Set$ together with a relation
$\leq \, : \brouwer \to \brouwer \to \Prop$.  The constructors for
$\brouwer$ are
\begin{alignat}{3}
& \bzero &:&\;  &&\brouwer \\
& \bsuc &:& && \brouwer \to \brouwer \\
& \blimit &:& && (\N \toincr{<} \brouwer) \to \brouwer \\
& \bbisim & : & && (f \, g : \N \toincr{<} \brouwer) \to  f \approx^{\leq} g \to \blimit \, f = \blimit \, g,
\end{alignat}
where we write $x < y$ for $\bsuc \, x \leq y$ in the type of
$\blimit$. Simulations thus use $\leq$ and the \emph{increasing}
predicate uses $<$, as one would expect.  The truncation constructor,
ensuring that $\brouwer$ is a set, is kept implicit in the paper (but
is explicit in the Agda formalisation).

The mutually defined relation $\leq$ is inductively defined by the following constructors, where each constructor
is implicitly quantified over the variables $x, y, z : \brouwer$ and
$f : \N \toincr{<} \brouwer$ that it contains:
\begin{alignat}{3}
&\lzero &:& &\quad& 
    \bzero \leq x \\
&\ltrans &:& && 
    x \leq y \to y \leq z \to x \leq z\\
&\lsuccmono &:& && 
    x \leq y \to \bsuc \, x \leq \bsuc \, y \\
&\lcocone &:& && 
    (k : \N) \to x \leq f(k) \to x \leq \blimit \, f \\
&\llimiting &:& && 
    (\forall k. f(k) \leq x) \to \blimit \, f \leq x
\end{alignat}
The truncation constructor, which ensures that $x \leq y$ is a
proposition, is again kept implicit.

We hope that the constructors of $\brouwer$ and $\leq$ are
self-explanatory.  $\lcocone$ ensures that $\blimit \, f$ is indeed an
upper bound of $f$, and $\llimiting$ witnesses that it is the
\emph{least} upper bound or, from a categorical point of view, the
(co)limit of $f$.

By restricting to limits of strictly increasing sequences, we can avoid
the representation of zero or successor ordinals as limits (as otherwise e.g.\
$a = \blimit \, (\lambda i . a)$). If one wishes to drop this
restriction, it will be necessary to strengthen the $\bbisim$ constructor to
witness antisymmetry --- however, we found that version of $\brouwer$
significantly harder to work with; see \cref{subsec:alt-def-of-Brouwer} for a short discussion.
Another question is whether adding the constructor $\ltrans$ explicitly is necessary since,
even without including $\ltrans$ in the construction, it might be possible to
derive transitivity of $\leq$ anyway.
We do not know the answer to that question.

\subsection{Transitive, Extensional and Well-Founded Orders}
\label{subsec:bookords}

The third notion of ordinals that we consider is the one studied in
the HoTT book~\cite{hott-book}.  This is the notion which is closest
to the classical definition of an ordinal as a set with a
well-founded, trichotomous, and transitive
order, 
without a concrete representation.  Requiring trichotomy leads to a
notion that makes many constructions impossible in a setting where the
law of excluded middle is not assumed.  Therefore, when working
constructively, it is better to replace the axiom of trichotomy by
\emph{extensionality}, stating that any two elements of $X$ with the
same predecessors are equal.

Concretely, an ordinal in the sense of the HoTT book~\cite[Def~10.3.17]{hott-book}
is a type%
\footnote{Note that the HoTT book~\cite[Def~10.3.17]{hott-book} asks for $X$ to be a
  set, but Escard\'o~\cite{escardo:agda-ordinals} proved that this follows from the rest of the definition, and we
  therefore drop this requirement.}  $X$ together with a relation
$\prec \, : X \to X \to \Prop$ which is \emph{transitive},
\emph{extensional}, and \emph{well-founded} (every element is accessible, where
accessibility is the least relation such that $x$ is accessible if
every predecessor of $x$ is accessible) --- we will recall the
precise definitions in \cref{sec:axiomaticapproach}.  We write
$\bookord$ for the type of ordinals in this sense.
Note the shift of universes that happens here: the type $\bookord_i$
of ordinals with $X : \UU_i$ is itself in $\UU_{i+1}$. We are mostly
interested in $\bookord_0$, but note that $\bookord_0$ lives in
$\UU_1$, while $\cnf$ and $\brouwer$ both live in $\UU_0$.

We also have a relation on $\bookord$ itself.  Following the HoTT book~\cite[Def~10.3.11 and Cor~10.3.13]{hott-book}, a \emph{simulation} between
ordinals $(X,\prec_X)$ and $(Y,\prec_Y)$ is a function $f : X \to Y$
such that:
\begin{enumerate}[(a)]
\item $f$ is monotone:
  $(x_1 \prec_X x_2) \to (f\, x_1 \prec_Y f\, x_2)$; and
\item for all $x : X$ and $y : Y$, if $y \prec_Y f\, x$, then we have
  an $x_0 \prec_X x$ such that $f\, x_0 =
  y$. \label{item:simu-property}
\end{enumerate}
We write $X \leq Y$ for the type of simulations between $(X,\prec_X)$
and $(Y,\prec_Y)$.
Given an ordinal $(X,\prec)$ and $x:X$, the \emph{initial segment} of
elements below $x$ is given as
$X_{\slash x} \defeq \Sigma(y : X). y \prec x$.  Again following the HoTT book~\cite[Def~10.3.19]{hott-book}, a simulation $f : X \leq Y$ is \emph{bounded} if
we have $y : Y$ such that $f$ induces an equivalence
$X \simeq Y_{\slash y}$.  We write $X < Y$ for the type of bounded
simulations.  This completes the definition of $\bookord$ together
with type families $\leq$ and $<$.


\section{An Abstract Axiomatic Framework for Ordinals} \label{sec:axiomaticapproach}

Which properties do we expect a type of ordinals to have?
Compared to the previous section, we go up one level of abstraction and consider an
arbitrary set $A$ together with relations $<$ and $\leq$ valued in propositions:
\begin{alignat}{3}
&A && \colon && \Set \\
&(\_\!\!<\!\!\_) && \colon && A \to A \to \Prop \\
&(\_\!\!\leq\!\!\_) && \colon && A \to A \to \Prop.
\end{alignat}
The types $\cnf$, $\brouwer$, $\bookord$ with their relations are concrete implementations of such a triple $(A,<,\leq)$.%
\footnote{Note that we do not require $A$ to live in a specified universe, as $\bookord$ is larger than $\cnf$ or $\brouwer$.}
In the current section, we discuss the various constructions that a concrete implementation may or may not allow, and list the main results (see the ``summary boxes'' below and on the next pages); in \cref{sec:CNF,sec:HIIT,sec:bookords}, we discuss $\cnf$, $\brouwer$, and $\bookord$ respectively in detail, and prove the precise theorems.

For $\cnf$, the relation $\leq$ is the reflexive closure of $<$, but the analogous statement is not constructively provable for $\brouwer$ and $\bookord$.
In the current section, we make the following basic assumptions:
\begin{enumerate}[({A}1)]
	\item \label{item:assumption1}
	$<$ is transitive ($x < y \to y < z \to x < z$) and irreflexive $(\neg(x < x)$);
	\item \label{item:assumption2}
	$\leq$ is reflexive ($x\leq x$), transitive, and antisymmetric ($x \leq y \to y \leq x \to x = y$);
	\item \label{item:assumption3}
	we have $(<) \subseteq (\leq)$ and $(< \circ \leq) \subseteq (<)$.
      \end{enumerate}
On top of these assumptions, we can now consider additional properties that we would expect ordinals to have.
The third condition (A\ref{item:assumption3}) means that $(x < y) \to (x \leq y)$ and $(x < y) \to (y \leq z) \to (x < z)$.
We do not assume the ``symmetric'' variation
\begin{equation} \label{eq:leq-<-bad-direction}
(\leq \circ <) \subseteq (<),
\end{equation}
which is true for $\cnf$ and $\brouwer$, but only holds for $\bookord$ iff $\LEM$ holds.
This constructive failure is known and can be seen as motivation for \emph{plump} ordinals~\cite{shulman:plump, taylor:ordinals}.

Proving that $\leq$ for $\brouwer$ is antisymmetric is challenging because of the path constructors in the inductive-inductive definition of Brouwer trees.
Of course, this difficulty arises as a consequence of our chosen definition for $\brouwer$, and other definitions would make antisymmetry easy; but unsurprisingly, such alternative definitions simply shift the difficulties to other places, see \cref{subsec:alt-def-of-Brouwer}.

In \cref{sec:CNF,sec:HIIT,sec:bookords}, we will prove in detail which of the properties discussed here hold for $\cnf$, $\brouwer$, and $\bookord$.
In the current section, we very briefly summarise these results in boxes such as the one below.
While some of the stated properties are original results of the current paper, others are known and stated for comparison only; the references included in the boxes lead to the precise theorems and proofs or citations.

\begin{summary}
	The assumptions (A\ref{item:assumption1}), (A\ref{item:assumption2}), and (A\ref{item:assumption3}) are satisfied for $\cnf$, $\brouwer$, and $\bookord$.
    The property $(\leq \circ <) \subseteq (<)$ holds for $\cnf$ and $\brouwer$, but is equivalent to $\LEM$ for $\bookord$.
	
	\summarySpace
	\useBoxCref
	
	\begin{flushright}
		\small
		Precise statements: \cref{thm:CNF-satisfies-general-notions} ($\cnf$); \cref{cor:brouwer-satisfies-assumptions} ($\brouwer$); \cref{ord-wrong-transitivity,cor:bookord-satisfies-assumptions} ($\bookord$).
	\end{flushright}
\end{summary}

For the rest of \cref{sec:axiomaticapproach}, we assume that $(A, <, \leq)$ is given and satisfies the conditions above.

\subsection{Extensionality and Well-Foundedness}\label{subsec:ext-wf}


Following the HoTT book~\cite[Def~10.3.9]{hott-book}, we call a relation $\prec$ \emph{extensional} if, for all $a, b : A$, we have 
\begin{equation}
(\forall c. c \prec a \leftrightarrow c \prec b) \to a = b.
\end{equation}
Assumption (A\ref{item:assumption2}) implies that $\leq$ is extensional: given $a$ and $b$ such that $c \leq a \leftrightarrow c \leq b$ for every $c$, we have $a \leq a$ by reflexivity, and hence $a \leq b$ by assumption. Similarly, we get $b \leq a$, and hence $a = b$ by antisymmetry.
In contrast, extensionality is not guaranteed for $<$, but we will show that it holds for our three instances $\cnf$, $\brouwer$, and $\bookord$.
This is non-trivial in the case of $\brouwer$ --- note that it fails for the ``naive'' version of $\brouwer$, where the path constructor $\bbisim$ is missing.


We use the inductive definition of accessibility and well-foundedness (with respect to $<$) by Aczel~\cite{aczelinductive}.
Concretely, the type family
$\acc : A \to \UU$ is inductively defined by the constructor
\begin{equation}
\mathit{access} : (a : A) \to ((b : A) \to b < a \to \acc(b)) \to \acc(a).
\end{equation}
An element $a : A$ is called \emph{accessible} if $\acc(a)$, and $<$ is \emph{well-founded} if all elements of $A$ are accessible. It is well known that the following induction principle can be derived from the inductive presentation~\cite{hott-book}:

\begin{lemma}[\flinkFull{https://cj-xu.github.io/agda/type-theoretic-approaches-to-ordinals/index.html\#2664}, transfinite induction]
  \label{lm:wf:ti}
    Let $<$ be well-founded and $P : A \to \UU$ be a type family such that
    $\forall a. (\forall b<a.P(b)) \to P(a)$.
    Then, it follows that $\forall a. P(a)$.
    \qed
\end{lemma}

In all our use cases in this paper, $P$ will be a mere property (i.e.\ a family of propositions), although the induction principle is valid even without this assumption.

As a standard sample application, we show that the classical formulation of well-foundedness is a consequence:
\begin{lemma}[\flinkFull{https://cj-xu.github.io/agda/type-theoretic-approaches-to-ordinals/index.html\#2829}]
  \label{lem:no-infinite-sequence}
    If $<$ is well-founded, then there is no infinite decreasing sequence:
    \begin{equation}
    \neg \left( \Sigma (f : \N \to A). (i : \N) \to f(i+1) < f(i) \right).
    \end{equation}
    In particular, there is no cycle $a_0 < a_1 < \ldots < a_n < a_0$.
    For $n \equiv 0$, this says that $<$ is irreflexive.
\end{lemma}
\begin{proof}
    We apply \cref{lm:wf:ti} with the property $P$ given by
    \begin{equation}
    P(a) \; \defeq \; \neg \Sigma (f : \N \to A). (f\, 0 = a) \times \left((i : \N) \to f(i+1) < f(i)\right).
    \end{equation}
    To show the induction step, assume a sequence $f$ with $f(0) = a$ is given.
    Then, the sequence $\lambda i. f(i+1)$ gives a contradiction by the induction hypothesis.
\end{proof}

From the global assumptions that $A$ is a set and $<$ is irreflexive as well as valued in propositions, we get that $x < y$ and $x = y$ are mutually exclusive propositions.
Therefore, we get the following observation for the reflexive closure:

\begin{lemma}[\flinkFull{https://cj-xu.github.io/agda/type-theoretic-approaches-to-ordinals/index.html\#3019}]
    The reflexive closure of $<$, given by $(x < y) \dissum (x = y)$, is valued in propositions. \qed
\end{lemma}


\begin{summary}
	For each of $\cnf$, $\brouwer$, and $\bookord$, the relation $<$ is extensional and well-founded.
	
\summarySpace
\useBoxCref

\begin{flushright}
	\small
	Precise statements: 
	\cref{thm:CNF-satisfies-general-notions,thm:cnf-wellfounded} ($\cnf$); 
	\cref{thm:brouwer-wellfounded,thm:brouwer-extensional} ($\brouwer$);
	\cref{ord-ext-wf-transitive} ($\bookord$).
\end{flushright}

\end{summary}

Note that the results stated so far in particular mean that $\cnf$ and $\brouwer$ can be seen as elements of $\bookord$ themselves.

\subsection{Classification as Zero, a Successor, or a Limit}\label{subsec:class}

All standard formulations of ordinals allow us to determine a minimal ordinal \emph{zero} and (constructively) calculate the \emph{successor} of an ordinal, but only some allows us to also calculate the \emph{supremum} or \emph{limit} of a family of ordinals.

\subsubsection{Zero, Successors, and Suprema}\label{subsubsec:zero-succ-sup}

Let $a$ be an element of $A$.
It is \emph{zero}, or \emph{bottom}, if it is at least as small as any other element
\begin{equation}
  \label{eq:iszero}
\iszero(a) \defeq \forall b. a \leq b,
\end{equation}
and we say that the triple $(A,<,\leq)$ \emph{has a zero} if we have an inhabitant of the type $\Sigma(z : A). \iszero(z)$. Both the types ``being a zero'' and ``having a zero'' are propositions.

We say that $a$ is a \emph{successor} of $b$ if it is the least element strictly greater\footnote{Note that $>$ is the obvious symmetric notation for $<$; it is \emph{not} a newly assumed relation.} than $b$:
\begin{equation}
  (\isupsucof a b) \defeq (a > b) \times \forall x > b. a \leq x.
\end{equation}
We say that $(A,<,\leq)$ \emph{has successors} if
there is a function $s : A \to A$ which calculates successors, i.e.\ such that $\forall b. \issucof{s(b)}{b}$. ``Calculating successors'' and ``having successors'' are propositional properties, i.e.\ if a function that calculates successors exists, then it is unique. The following statement is simple but useful. Its proof uses assumption (A\ref{item:assumption3}).
\begin{lemma}[\flinkFull{https://cj-xu.github.io/agda/type-theoretic-approaches-to-ordinals/index.html\#3494}]
  \label{thm:calc-succ-characterisation}
  A function $s : A \to A$ calculates successors if and only if $\forall b\, x. (b < x) \leftrightarrow (s\, b \leq x)$.
  \qed
\end{lemma}

Dual to ``$a$ is the least element strictly greater than $b$'' is the 
statement that ``$b$ is the greatest element strictly below $a$'', in which case it is natural to call $b$ the \emph{predecessor} of $a$.
If $a$ is the successor of $b$ and $b$ the predecessor of $a$, then we call $a$ the \emph{strong successor} of $b$:
\begin{equation}
(\isstrongsucof{a}{b}) \defeq \issucof{a}{b} \times \forall x < a. x \leq b.
\end{equation}
We say that $A$ \emph{has strong successors} if there is $s : A \to A$ which calculates strong successors, i.e.\ such that $\forall b. \isstrongsucof{s(b)}{b}$.
The additional information contained in a strong successor plays an important role in our technical development. 

%

Finally, we consider the \emph{suprema} or, synonymously, \emph{least upper bounds} of families of ordinals.
Given a type $X$ and $f : X \to A$, an element $a : A$ is the supremum of $f$ if it is at least as large as every $f(x)$, and if any $b$ with this property is at least as large as $a$:
\begin{equation}\label{eq:is-sup-of}
(\isupsupof a f) \defeq (\forall x. f(x) \leq a) \times (\forall b. (\forall x. f(x) \leq b) \to a \leq b). 
\end{equation}
We say that $(A,<,\leq)$ \emph{has suprema of $X$-indexed families} if there is a function $\sqcup : (X \to A) \to A$ which calculates suprema, i.e.\ such that $(f : X \to A) \to \issupof{(\sqcup f)}{f}$. Note that the supremum is unique if it exists, i.e.\ the type of suprema is propositional for a given pair $(X,f)$.
Both the properties ``calculating suprema'' and ``having suprema'' are propositions.
If $(A,<,\leq)$ has suprema of $\Bool$-indexed families, we also say that $A$ has \emph{binary joins}.
Unless explicitly specified, in the following we will consider suprema of $\Nat$-indexed families only.

Instead of considering functions $f$ without further structure, we can ask for $f$ to be a morphism of (partial) orders.
Of particular interest is the case of \emph{$\leq$-monotone} $\N$-indexed sequences, i.e.\ functions $f : \N \to A$ such that $f_n \leq f_{n+1}$.
We also consider strictly increasing $\N$-indexed sequences satisfying the condition $f_n < f_{n+1}$.
Note that every $a : A$ is trivially the supremum of the sequence constantly $a$, and therefore, ``being a supremum'' does not describe the usual notion of \emph{limit ordinals}.
One might consider $a$ a \emph{proper} supremum of $f$ if $a$ is pointwise strictly above $f$, i.e.\ $\forall i. f_i < a$.
This is automatically guaranteed for strictly increasing $\N$-sequences,
and in this case, we call $a$ the \emph{limit} of $f$:
\begin{align}
& \islimof \blank \blank : A \to (\N \toincr{<} A) \to \UU \\
& \islimof{a}{(f,q)} \defeq \issupof{a}{f}.
\end{align}
We say that $A$ \emph{has limits} if there is a function $\mathsf{limit} : (\N \toincr{<} A) \to A$ that calculates limits of strictly increasing $\N$-sequences.
Note that $\cnf$ cannot have limits since one can construct a sequence (see \cref{thm:cnf-below-eps0}) which comes arbitrarily close to $\varepsilon_0$.
The question becomes more interesting if we consider \emph{bounded} sequences, i.e.\ sequences $f$ with $b : \cnf$ such that $f_i < b$ for all $i : \N$.

\begin{summary}
All our notions of ordinals have zeroes and strong successors.
$\bookord$ is the only considered notion that constructively has suprema of arbitrary small families --- this was proven by de Jong and Escard\'o~\cite{tom_smalltypes}.
$\cnf$ only has suprema of finite families, which $\brouwer$ does not have constructively; on the other hand, $\brouwer$ has suprema of strictly increasing $\N$-indexed sequences.
Monotonicity of the successor function holds for $\cnf$ and $\brouwer$, but not constructively for $\bookord$.

\summarySpace
\useBoxCref

\begin{flushright}
	\small
	Precise statements: 
	\cref{lem:cnf-zero-suc,thm:cnf-sups-lims-proof} ($\cnf$); 
	\cref{thm:brouwer-has-it-all,cor:brw-suc-mono,thm:brw-lpo-to-join-to-wlpo} ($\brouwer$); 
	\cref{thm:bookord-zersuclim,thm:ord-everything-undecidable} ($\bookord$).
\end{flushright}
\end{summary}

\subsubsection{Classifiability}

For classical set-theoretic ordinals, every ordinal is either zero, a successor, or a limit.
We say that a notion of ordinals which allows this has classification.
This is very useful, as many theorems that start with ``for every ordinal'' have proofs that consider the three cases separately.
In the same way as not all definitions of ordinals make it possible to calculate limits, only some formulations make it possible to constructively classify any given ordinal.
We already defined what it means to be zero in \eqref{eq:iszero}.
We now also define what it means for $a : A$ to be a strong successor or a limit of a strictly increasing $\N$-indexed sequence:
\begin{align}
  &\isstrongsuc(a) \defeq \Sigma(b:A).  (\isstrongsucof a b)\\
  &\islim(a) \defeq \exists f : \N \toincr{<} A. \, \islimof{a}{f}.
\end{align}
In addition, we say that $a$ is a general limit if it is the supremum of the set of all elements below $a$ (and there exists at least one such element).
As in \cref{subsec:bookords}, we write $A_{\slash a} \defeq \Sigma(b : A). b < A$ for this type.
We have the first projection $\fstproj : A_{\slash a} \to A$ and define:%
\begin{equation}\label{eq:being-general-limit}
	\isgenerallim(a) \defeq \proptrunc{A_{\slash a}} \times (\issupof{a}{\fstproj}).
\end{equation}

\begin{remark}[\flinkFull{https://cj-xu.github.io/agda/type-theoretic-approaches-to-ordinals/index.html\#3624}]
  One can also consider to define $a$ to be a general limit if $a$ is the supremum of \emph{some} small and inhabited family of elements below $a$, i.e., if there exists a small inhabited type $X$ and $f : X \to A$ such that $f(x) < a$ for every $x : X$, and $\issupof{a}{f}$. Note that every general limit in this sense is also a general limit in the sense of \eqref{eq:being-general-limit}, and if $A$ is small, then the two notions are equivalent. However, the type of transitive, extensional, and well-founded orders in particular is not small, and so this   definition is different from \eqref{eq:being-general-limit} for $\bookord$, as $\fstproj : \bookord_{\slash X} \to \bookord$ is a \emph{large} family.
We do not know whether the relevant part of \cref{thm:ord-everything-undecidable} holds for this notion of general limits, i.e., whether $\LEM$ implies that every $X$ is either zero or a successor or the limit of a small family.
\end{remark}

All of $\iszero(a)$, $\isstrongsuc(a)$, $\islim(a)$, and $\isgenerallim(a)$ are propositions. This is true even though $\isstrongsuc(a)$ is defined without a propositional truncation because,
if $a$ is the strong successor of both $b$ and $b'$, we have $b \leq b'$ and $b' \leq b$, implying $b = b'$ by antisymmetry.

\begin{lemma}[\flinkFull{https://cj-xu.github.io/agda/type-theoretic-approaches-to-ordinals/index.html\#3974}]
  \label{lem:only-one-out-of-three}
    Any $a:A$ can be at most one out of \{zero, strong successor, limit\}, and in a unique way.
    In other words, the type $\iszero(a) \dissum \isstrongsuc(a) \dissum \isgenerallim(a)$ is a proposition.
	Similarly, the type $\iszero(a) \dissum \isstrongsuc(a) \dissum \islim(a)$ is a proposition.
\end{lemma}
\begin{proof}
     As mentioned above, all considered summands are propositions.
	 What we have to check is that they mutually exclude each other.
     Note that $\islim(a)$ implies $\isgenerallim(a)$, which means it suffices to consider the case of $\iszero(a) \dissum \isstrongsuc(a) \dissum \isgenerallim(a)$.

     Since the goal is a proposition, we can assume that we are given $b$ and $b_0<a$ in the successor and limit case.
     Assume that $a$ is zero and the successor of $b$. This implies $b < a \leq b$ and thus $b < b$, contradicting irreflexivity.
     If $a$ is zero and the limit of $\fstproj : A_{\slash a} \to A$, the same argument (with $b$ replaced by $b_0$) applies.
     Finally, assume that $a$ is the strong successor of $b$ and the limit of $\fstproj$. These assumptions show that $b$ is an upper bound of $\fstproj$, thus we get $a \leq b$. Together with $b < a$, this gives the contradiction $b < b$ as above.
\end{proof}

We say that an element of $A$ is \emph{weakly classifiable}
if it is zero or a strong successor or a general limit, and \emph{classifiable}
if it is zero or a strong successor or a limit of a strictly increasing $\N$-indexed sequence.
We say $(A,<,\leq)$ \emph{has (weak) classification} if every element of $A$ is (weakly) classifiable.%
\footnote{As the terminology suggests, we focus on limits of increasing sequences and classification, while weak classification only plays a very minor role. The reason is that none of our notions of ordinals has weak classification without having classification, and the latter is easier to work with.}
By \cref{lem:only-one-out-of-three}, $(A, <, \leq)$ has classification exactly if the type $\iszero(a) \dissum \isstrongsuc(a) \dissum \islim(a)$
is contractible, in the jargon of homotopy type theory.


Classifiability corresponds to a case distinction, but the useful
principle from classical ordinal theory is the related induction principle:

\begin{definition}[classifiability induction]
    \label{def:CFI}
    We say that $(A,<,\leq)$ satisfies the principle of \emph{classifiability induction} if the following holds:
    For every family $P : A \to \Prop$ such that
    \begin{align}
    & \iszero(a) \to P(a) \label{eq:class-ind-zero} \\
    & (\isstrongsucof{a}{b}) \to P(b) \to P(a) \label{eq:class-ind-suc} \\
    & (\islimof a f) \to (\forall i. P(f_i)) \to P(a), \label{eq:class-ind-lim}
    \end{align}
    we have $\forall a. P(a)$.
\end{definition}

Note that classifiability induction does \emph{not} ask that there are functions that computes successors or limits.
The following is immediate:
\begin{corollary}[\flinkFull{https://cj-xu.github.io/agda/type-theoretic-approaches-to-ordinals/index.html\#4331}, of \cref{lem:only-one-out-of-three}] \label{thm:CFI-to-classification}
	If $(A,<,\leq)$ satisfies classifiability induction, then it has classification. \qed
\end{corollary}
For the reverse direction, we need a further assumption:
\begin{theorem}[\flinkFull{https://cj-xu.github.io/agda/type-theoretic-approaches-to-ordinals/index.html\#4475}]
    \label{thm:CFI}
    Assume $(A,<,\leq)$ has classification and satisfies the principle of transfinite induction.
    Then $(A,<,\leq)$ satisfies the principle of classifiability induction.
\end{theorem}
 \begin{proof}
     With the assumptions of the statement and \eqref{eq:class-ind-zero},\eqref{eq:class-ind-suc}, and \eqref{eq:class-ind-lim}, we need to show $\forall a.P(a)$.
     By transfinite induction, it suffices to show
     \begin{equation} \label{eq:tfi-ass}
     (\forall b<a. P(b)) \to P(a)
     \end{equation}
     for some fixed $a$.
     By classification, we can consider three cases.
     If $\iszero(a)$, then \eqref{eq:class-ind-zero} gives us $P(a)$, which shows \eqref{eq:tfi-ass} for that case.
     If $a$ is the strong successor of $b$, we use that the predecessor $b$ is one of the elements that 
     the assumption of \eqref{eq:tfi-ass} quantifies over; therefore, this is implied by 
     \eqref{eq:class-ind-suc}.
     Similarly, if $\islim(a)$, the assumption of \eqref{eq:tfi-ass} gives $\forall i. P(f_i)$, thus \eqref{eq:class-ind-lim} gives $P(a)$.
 \end{proof}

It is also standard in classical set theory that classifiability induction implies transfinite induction:
showing $P$ by transfinite induction corresponds to showing $\forall x<a. P(x)$ by classifiability induction.
In our setting, this would require strong additional assumptions, including the assumption that $(x \leq a)$ is equivalent to $(x < a) \dissum (x = a)$, i.e.\ that $\leq$ is the reflexive closure of $<$.
The standard proof works with several strong assumptions of this form, but we do not consider this interesting or useful, and concentrate on the results which work for the weaker assumptions (A\ref{item:assumption1}), (A\ref{item:assumption2}), (A\ref{item:assumption3}) that are satisfied for $\brouwer$ and $\bookord$.

\begin{summary}
	$\cnf$ and $\brouwer$ have classification and satisfy classifiability induction, while $\bookord$ does not, constructively.
    $\bookord$ has weak classification if an only if $\LEM$ holds.

\summarySpace
\useBoxCref

\begin{flushright}
	\small
	Precise statements:  
	\cref{cnf-has-classification} ($\cnf$); 
	\cref{thm:brw-class-CFI} ($\brouwer$); 
	\cref{thm:ord-everything-undecidable} ($\bookord$).
\end{flushright}
\end{summary}


\subsection{Arithmetic}
\label{subsec:abstract-arithmetic}

Using the predicates $\iszero(a)$, $\issucof a b$, and $\issupof{a}{f}$, we can define what it means for $(A,<,\leq)$ to have the standard arithmetic operations.
We still work under the assumptions declared at the beginning of the section --- in particular, we do not assume that e.g.\ limits can be calculated, which is important to make the theory applicable to $\cnf$.

\begin{definition}[having addition] \label{def:have-addition}
    We say that $(A,<,\leq)$ \emph{has addition}
    if there is a function $+ : A \to A \to A$ which satisfies the following properties:
    \begin{align}
    & \iszero(a) \to c + a = c \\
    & \issucof a b \to \issucof {d} {(c+b)} \to c + a = d \\
    & \islimof{a}{f} \to \issupof b {(\lambda i. c + f_i)} \to c + a = b  \label{eq:having-addition-3}
    \end{align}
    We say that $A$ \emph{has unique addition} if there is exactly one function $+$ with these properties.
\end{definition}

Note that \eqref{eq:having-addition-3} makes an assumption only for (strictly) \emph{increasing} sequences $f$; this suffices to define a well-behaved notion of addition, and it is not necessary to include a similar requirement for arbitrary sequences.
Since $(\lambda i. c + f_i)$ is a priori not necessarily increasing, the middle term of \eqref{eq:having-addition-3} has to talk about the supremum, not the limit.

Completely analogously to addition, we can formulate multiplication and exponentiation, again without assuming that successors or limits can be calculated:

\begin{definition}[having multiplication] \label{def:have-multiplication}
    Assuming that $A$ has addition $+$, we say that it \emph{has multiplication}
    if we have a function $\cdot : A \to A \to A$ that satisfies the following properties:
    \begin{align}
    & \iszero(a) \to c \cdot a = a \\
    & \issucof a b \to c \cdot a = c \cdot b + c \\
    & \islimof{a}{f} \to \issupof b {(\lambda i. c \cdot f_i)} \to c \cdot a = b
    \end{align}
    $A$ \emph{has unique multiplication} if it has unique addition and there is exactly one function $\cdot$ with the above properties.
\end{definition}

\begin{definition}[having exponentiation]
\label{def:have-exponentiation}
    Assume $A$ has addition $+$ and multiplication $\cdot$.
    We say that $A$ \emph{has exponentiation with base $c$}
    if we have a function $c^- : A \to A$ that satisfies the following properties:
    \begin{align}
    & \iszero(b) \to \issucof a b \to c^b = a \\
    & \issucof a b \to c^a = c^b \cdot c \\
    & \begin{multlined}
    \islimof{a}{f} \to \neg \iszero(c) \to {\issupof{b}{c^{f_i}} \to c^a = b}
    \end{multlined} \\
    & \islimof{a}{f} \to \iszero(c) \to c^a = c
    \end{align}
    $A$ \emph{has unique exponentiation with base $c$} if it has unique addition and multiplication, and if $c^-$ is unique.
\end{definition}

Let us now define subtraction, an operation that can be specified using addition.
Note that there is more than one canonical choice: Given numbers $a$ and $b$ with $a \leq b$, we can either require their difference $c$ to satisfy $a + c = b$ or $c + a = b$ (or even both),
but only the first option (\emph{left subtraction}) is in line with the specification for addition.%
\footnote{If $a$ is a limit, then $c + a$ cannot be a successor, and vice versa; in other words, requiring $c + a = b$ would imply that the difference between a limit and a successor cannot exist.}

\begin{definition}[having subtraction] \label{def:have-left-subtraction}
	We say that $(A,<,\leq)$ \emph{has subtraction}
	if it has addition $+$, and a function $- : (b : A) \to (a : A) \to (p : a \leq b) \to A$, written $b -_p a$, such that
	$a + (b -_p a) = b$.
	We say that $A$ \emph{has unique subtraction} if it has unique addition and there is exactly one function $-$ with these properties.
\end{definition}

Completely analogously, it would be possible to specify division and logarithm.
Such constructions are further discussed in \cref{sec:CNF} below in the context of Cantor normal forms, but play otherwise no role in this paper.


\begin{summary}
$\cnf$ uniquely has addition, multiplication, exponentiation with base $\omega$,
subtraction, and division.
%
	$\brouwer$ uniquely has addition, multiplication, and exponentiation. 
	Further, $\brouwer$ has subtraction (necessarily unique) if and only if $\LPO$ holds. 
	$\bookord$ has addition and multiplication, but subtraction is available if and only if $\LEM$ holds.

\summarySpace
\useBoxCref

\begin{flushright}
	\small
	Precise statements: 
	\cref{thm:cnf-arith,lem:cnf-sub-div,thm:cnf-arith-unique} ($\cnf$); 
	\cref{thm:brw-has-all-arithmetic-properties,thm:brw-subtraction-iff-lpo} ($\brouwer$);
	\cref{thm:ord-addition-multiplication,thm:ord-subtraction-iff-lem} ($\bookord$).
\end{flushright}

\end{summary}

\subsection{Constructive Aspects}\label{subsec:Decidability}

The main differences between our three versions of constructive ordinals are their varying degrees of decidability of certain properties.
As described in the introduction, we view Cantor normal forms as a notion where almost everything is decidable, while basically nothing is decidable for extensional well-founded orders, and Brouwer trees sit in the sweet spot in the middle.
In this section, we want to make this intuition precise.

We say that a proposition%
\footnote{While these definitions do not require $P$ to be a proposition but work for any type, we only use them for propositions.}
$P$ is \emph{decidable} if we have either $P$ or $\neg P$, i.e.
\begin{equation}
\Dec(P) \defeq P \dissum \neg P,
\end{equation}
and \emph{stable} if its double negation is as strong as $P$,
\begin{equation}
\Stable(P) \defeq \neg\neg P \to P.
\end{equation}
Of course, decidable propositions are always stable.

Given a set $A$, we can then ask whether equality is stable ($\forall (x,y : A). \Stable(x=y)$)
or even decidable ($\forall (x,y : A). \Dec(x=y)$).
Slightly weakening the properties we can, for a given $x_0 : A$, ask whether equality is locally stable ($\forall (y : A). \Stable(x_0=y)$) or decidable ($\forall (y : A). \Dec(x_0=y)$).
If the set comes with relations $<$, $\leq$, we can ask the same questions for these.
Moreover, we can ask whether $\leq$ \emph{splits},
\begin{equation}
  \label{eq:splits}
\Splits(A,<,\leq) \defeq \forall (x,y : A). (x \leq y) \to (x < y) \dissum (x = y).
\end{equation}

A relation $\leq$ is \emph{connex} if $(a \leq b) \dissum (b \leq a)$, and a relation $<$ is \emph{trichotomous} if $(a < b) \dissum (a = b) \dissum (b < a)$.
Note that these two properties are very strong; under mild additional assumptions, they imply most or all of the other discussed properties.
As an example, we have:
\begin{lemma}[\flinkFull{https://cj-xu.github.io/agda/type-theoretic-approaches-to-ordinals/index.html\#5131}]
  \label{lem:trich-splits}
	If $(A,<,\leq)$ satisfies the assumptions (A\ref{item:assumption1}) and $<$ is trichotomous, then assumption (A\ref{item:assumption3}) holds if and only if $\leq$ splits.
\end{lemma}
\begin{proof}
	We only show the direction ``only if''.
	Assume $x \leq y$. By trichotomy, we have $x < y$ or $x = y$ or $y < x$.
	In the first or second case, we are done. In the last case, (A\ref{item:assumption3}) implies $y<y$, contradicting (A\ref{item:assumption1}).
\end{proof}

When we work with concrete implementations of types of ordinals, it would of course be great to have a formulation that combines as many desirable properties as possible.
However, we cannot have certain properties at the same time, as demonstrated by the following no-go theorem:

\begin{theorem}[\flinkFull{https://cj-xu.github.io/agda/type-theoretic-approaches-to-ordinals/index.html\#5345}]
  \label{thm:no-go}
	Assume that $(A,<,\leq)$ has zero, successors, and limits of strictly increasing sequences.
	If $A$ has decidable equality,
	then $\WLPO$ holds.
\end{theorem}
\begin{proof}
	Zero $z$ and a successor function $s$ allow us to define a canonical strictly increasing function $\iota : \N \to A$.
	Let us write $\omega$ for the limit of this sequence.

	Let $t : \N \to \Bool$ be a binary sequence.
	We construct a sequence 
	\begin{equation}\label{eq:jumping-sequence-construction-only-one-jump-abstract}
	\jump t : \N \to A
	\end{equation} by
	\begin{align}
	& \jump t \, 0 \defeq z \\
	& \jump t \, (n+1) \defeq
	\begin{cases}
	\omega & \textit{if $n$ is minimal such that $t_n = \btt$}\\
	s (\jump t \, n) & \textit{else.}
	\end{cases}
	\end{align}
	We call $\jump t$ the \emph{jumping sequence} of $t$ as it ``jumps'' as soon as a $\btt$ is discovered in the sequence $t$.
	It is easy to see that $\jump t$ is strictly increasing. By assumption, it thus has a limit $j : A$.
	
	We claim that $j = \omega$ if and only if $\forall i. t_i = \bff$.
	If $j = \omega$, then $t_i = \btt$ leads to a contradiction for any $i$, thus we have $\forall i. t_i = \bff$.
	Vice versa, if $\forall i. t_i = \bff$, then $j = \omega$ by construction.
	
	Therefore, if the equality $j = \omega$ is decidable, then so is the property $\forall i. t_i = \bff$.
\end{proof}	

\cref{thm:no-go} means that a ``perfectly convenient'' implementation of ordinals cannot exist in a constructive world without $\WLPO$, and any implementation with zero and successors will have to sacrifice either decidable equality or limits.
In this paper, the three implementations demonstrating these choices are $\cnf$, $\brouwer$, and $\bookord$.

\begin{summary}
	Everything is decidable for $\cnf$ (as long as no infinite families of ordinals are involved), $\leq$ splits and is connex, and $<$ is trichotomous.
	The situation is very different for $\bookord$, where most properties that can be formulated using the above concepts imply or are equivalent to $\LEM$.
	
	\quad
	$\brouwer$ is the most interesting case: Many of the discussed properties are equivalent to $\LPO$, including decidability of the relations, splitting of $\leq$, 
	and trichotomy.
	At the same time, it is decidable whether $x : \brouwer$ is finite, and equalities/inequalities are decidable on the subtype of finite Brouwer tree ordinals.
	Local equality at $\omega$ is decidable if and only if $\WLPO$ holds, but local equality at $\omega \cdot 2$ 
	is already equivalent to $\LPO$.
	While local equality at $\omega$ is stable, local equality at $\omega \cdot 2$ implies $\MP$.

\summarySpace
\useBoxCref

\begin{flushright}
	\small
	Precise statements:
	\cref{thm:CNF-satisfies-general-notions} ($\cnf$); 
	\cref{thm:brw-decidable,thm:brw-LPO,thm:brw-dec-equ,thm:brw-dec-equ-omega-times-n,thm:brw-stable-equ,thm:brw-lpo-iff-trich-and-split} ($\brouwer$);  
	\cref{thm:ord-everything-undecidable,thm:ord-split-lem} ($\bookord$).
\end{flushright}

\end{summary}

\section{Cantor Normal Forms} \label{sec:CNF}

Ordinal notation systems based on Cantor normal forms (with or without coefficients) have been widely studied~\cite{Berger01a,CC:ord:coq,Dershowitz_superleaves,DR:ord:list,grimm:ord:coq,NFXG:three:ord,shinkarov-agda,takeuti:book}. In this section, we recall the well-known results of Cantor normal forms, adapted for our chosen representation $\cnf$ defined in Section~\ref{subsec:CNF}. We additionally prove that the arithmetic operations on $\cnf$ are uniquely correct with respect to our axiomatisation (Theorems~\ref{thm:cnf-arith} and~\ref{thm:cnf-arith-unique}), which has not been verified for Cantor normal forms previously, as far as we know.

As mentioned above, $\cnf$ provides a decidable formulation of ordinals in the following sense.

\begin{theorem}[\flinkFull{https://cj-xu.github.io/agda/type-theoretic-approaches-to-ordinals/index.html\#Theorem-18}]
  \label{thm:CNF-satisfies-general-notions}
	$\cnf$ is a set with decidable equality.
	The relations $<$ and $\leq$ are valued in propositions, decidable, extensional, and transitive.
	In addition, $<$ is irreflexive and trichotomous, while $\leq$ is reflexive, antisymmetric, connex, and splits. If $x \leq y$ and $y < z$, then $x < z$ follows.
\end{theorem}
\begin{proof}
	Most properties are proved by induction on the arguments. We prove the trichotomy property as an example. It is trivial when either argument is zero. Given $\tom a b$ and $\tom c d$, by the induction hypothesis we have $(a < c) \dissum (a = c) \dissum (c < a)$ correspondingly. For the first and last cases, we have $\tom a b < \tom c d$ and $\tom c d < \tom a b$. For the middle case, the induction hypothesis on $b$ and $d$ gives the desired result.
\end{proof}
Using \cref{lem:trich-splits}, the above theorem shows that $(\cnf, <, \leq)$ is an instantiation of the triple $(A, <, \leq)$ considered in \cref{sec:axiomaticapproach}.

%
We recall the following well-foundedness result of CNFs, which can be found in Nordvall Forsberg and Xu~\cite[Thm~5.1]{NFXG:three:ord} and in Grimm~\cite[\S2.3]{grimm:ord:coq}.
\begin{theorem}[\flinkFull{https://cj-xu.github.io/agda/type-theoretic-approaches-to-ordinals/index.html\#Theorem-19}]
  \label{thm:cnf-wellfounded}
The relation $<$ is well-founded. \qed
\end{theorem}

By~\cref{lm:wf:ti}, we obtain transfinite induction for CNFs. We next move on to arithmetic on CNF, which is defined using decidability of $<$.

\begin{definition}[Addition and multiplication of CNFs]
    We define addition and multiplication as follows.\footnote{Caveat: $\fatplus$ is a notation for the tree constructor, while $+$ is an operation that we define. We use parenthesis so that all operations can be read with the usual operator precedence.}
    \begin{alignat}{3}
    \tz + b & \defeq b \\
    a + \tz & \defeq a \\
    (\tom{a}{c}) + (\tom{b}{d}) & \defeq
    \begin{cases}
    \tom{b}{d} & \text{if $a<b$} \\
    \tom{a}{(c+\tom{b}{d})} & \text{otherwise}
    \end{cases} \\[4pt]
    \tz \cdot b & \defeq \tz \\
    a \cdot \tz & \defeq \tz \\
    a \cdot (\tom{\tz}{d}) & \defeq a + a \cdot d \\
    (\tom{a}{c}) \cdot (\tom{b}{d}) & \defeq (\tom{a+b}{\tz}) + (\tom{a}{c}) \cdot d \quad\text{if $b \not= \tz$}
    \end{alignat}
\end{definition}

The above operations are well-defined and have the expected ordinal arithmetic properties:
\begin{lemma}[\flinkFull{https://cj-xu.github.io/agda/type-theoretic-approaches-to-ordinals/index.html\#Lemma-21}]
\label{lem:cnf-arithmetic-properties-add-mul}
If $a,b$ are CNFs, then so are $a+b$ and $a \cdot b$. Both operations are associative and strictly increasing in the right argument. Moreover, $(\cdot)$ is distributive on the left, i.e., $a \cdot (b + c) = a \cdot b + a \cdot c$. \qed
\end{lemma}

Recall that we write $\tone$ as abbreviation for $\tom{\tz}{\tz}$.
It is immediate to check that:
\begin{lemma}[\flinkFull{https://cj-xu.github.io/agda/type-theoretic-approaches-to-ordinals/index.html\#Lemma-22}]
  \label{lem:cnf-zero-suc}
	$\tz$ is a zero (in the sense of \eqref{eq:iszero}), and $\lambda a. a + \tone$ gives strong successors that are $<$- and $\leq$-monotone. \qed
\end{lemma}

We also have:
\begin{definition}[Exponentiation with base $\omega$]
We define the CNF $\omega$ by $\omega \defeq \tom{\tone}{\tz}$ and the exponentiation $\omega^a$ of CNF $a$ with base $\omega$ by $\omega^a \defeq \tom{a}{\tz}$.
\end{definition}


It is easy to show that $\omega^{(-)}$ is exponentiation with base $\omega$ in the sense of Definition~\ref{def:have-exponentiation}. To show that $(+)$ is addition and $(\cdot)$ is multiplication in the sense of Definitions~\ref{def:have-addition} and~\ref{def:have-multiplication}, we need their inverse operations subtraction and division.

\begin{lemma}[\flinkFull{https://cj-xu.github.io/agda/type-theoretic-approaches-to-ordinals/index.html\#Lemma-24}] 
\label{lem:cnf-sub-div}
For all CNFs $a,b$,
\begin{enumerate}[(i)]
\item \label{item:sub} if $a \leq b$, then there is a CNF $c$ such that $a+c=b$ and thus we denote $c$ by $b-a$;
\item \label{item:div} if $b > \tz$, then there are CNFs $c$ and $d$ such that $a=b \cdot c + d$ and $d<b$. \qedhere
\end{enumerate}
\end{lemma}
\begin{proof}
For \eqref{item:sub}, we define subtraction $(-)$ as follows:
\begin{alignat}{3}
    \tz - b & \defeq \tz \\
    a - \tz & \defeq a \\
    (\tom{a}{c}) - (\tom{b}{d}) & \defeq
    \begin{cases}
    \tz & \text{if $a<b$} \\
    c-d & \text{if $a=b$} \\
    \tom{a}{c} & \text{if $a>b$}.
    \end{cases}
\end{alignat}
See our formalisation for the proof of correctness. The proof of \eqref{item:div} consists of the following cases:
\begin{itemize}
\item If $a<b$, then we take $c \defeq \tz$ and $d \defeq a$.
\item If $a=b$, then we take $c \defeq \tone$ and $d \defeq \tz$.
\item If $a>b$, then there two possibilities:
\begin{itemize}
\item $a=\tom u u'$ and $b=\tom v v'$ with $u>v$. By the induction hypothesis on $u'$ and $b$, we have $c'$ and $d$ such that $u'=b \cdot c' + d$ and $d<b$. We take $c \defeq \tom{(u-v)}{c'}$ and then have $a = \tom u u' = \omega^{v+(u-v)} + u' = b \cdot \omega^{(u-v)} + b \cdot c' + d = b \cdot c + d$.
\item $a=\tom u u'$ and $b=\tom u v'$ with $u'>v'$. By the induction hypothesis on $u'-v'$ and $b$, we have $c'$ and $d$ such that $u'-v' = b \cdot c' + d$ and $d<b$. We take $c \defeq c' + \tone$ and then have $a = \tom u u' = \omega^u + v' + (u' - v') = b + b \cdot c' + d = b \cdot c + d$.
\end{itemize}
\end{itemize}
The above defines the (Euclidean) division of CNFs.
\end{proof}

\begin{theorem}[\flinkFull{https://cj-xu.github.io/agda/type-theoretic-approaches-to-ordinals/index.html\#Theorem-25}]
\label{thm:cnf-arith}
$\cnf$ has addition $(+)$, multiplication $(\cdot)$ and exponentiation $\omega^{(-)}$ with base $\omega$.
\end{theorem}
\begin{proof}
We show the limit case for $(+)$, and refer to our formalisation for the rest.  Suppose $a$ is the limit of $f$. The goal is to show that $c+a$ is the supremum (and thus the limit) of $\lambda i. c+fi$. We know $c+fi \leq c+a$ for all $i$ because $(+)$ is increasing in the right argument (\cref{lem:cnf-arithmetic-properties-add-mul}). It remains to prove that if $c+fi \leq x$ for all $i$ then $c+a \leq x$. Thanks to Lemma~\ref{lem:cnf-sub-div}\eqref{item:sub}, we have $fi \leq x-c$ for all $i$ and thus $a \leq x-c$ because $a$ is the limit of $f$. Therefore, we have $c + a \leq c + (x - c) = x$.
\end{proof}

We conjecture that $\cnf$ has exponentiation with \emph{arbitrary} base.\footnote{The formalised proof is work in progress at the time of submission of this paper.} Specifically, we have constructed an operation $(-)^{(-)}$ and attempted to show a \emph{logarithm} lemma: for any CNFs $a > 0$ and $b > 1$, there are CNFs $x$, $y$ and $z$ such that $a = b^x \cdot y + z$ and $0<y<b$ and $z < b^x$.

All the arithmetic operations of CNFs are unique. An easy way to prove this fact is to use classifiability induction (Definition~\ref{def:CFI}) which we obtain as follows --- note that we can classify a CNF as a limit, even if we cannot compute limits of CNFs.

\begin{lemma}[\flinkFull{https://cj-xu.github.io/agda/type-theoretic-approaches-to-ordinals/index.html\#Lemma-26}]
  \label{lem:cnf:limit}
If a CNF is neither zero nor a successor, then it is a limit.
\end{lemma}
\begin{proof}
If a CNF $x$ is neither zero nor a successor, then $x = \tom a \tz$ with $a>0$ or $x = \tom a b$ where $b > 0$ is not a successor. There are three possible cases, for each of which we construct a strictly increasing sequence $s : \N \to \cnf$ whose limit is $x$:
\begin{enumerate}[(i)]
\item If $x = \tom a \tz$ and $a=c+1$, we define $s_i \defeq (\tom c 0) \cdot \eta i$ where $\eta : \N \to \cnf$ embeds natural numbers to CNFs. 
\item If $x = \tom a \tz$ and $a$ is not a successor, the induction hypothesis on $a$ gives a sequence $r$, and then we define $s_i \defeq \tom {r_i} \tz$. 
\item If $x = \tom a b$ and $b>0$ is not a successor, the induction hypothesis on $b$ gives a sequence $r$, and then we define $s_i \defeq \tom a {r_i}$. 
\end{enumerate}
The sequence $s$ is known as the \emph{fundamental sequence} of the CNF $x$.
\end{proof}

The construction of fundamental sequences for limit CNFs is standard and well known. For example, Grimm has developed it in Coq~\cite[\S2.5]{grimm:ord:coq}.

\begin{theorem}[\flinkFull{https://cj-xu.github.io/agda/type-theoretic-approaches-to-ordinals/index.html\#Theorem-27}]
  \label{cnf-has-classification}
	$\cnf$ has classification and satisfies classifiability induction.
\end{theorem}
\begin{proof}
	Since $\cnf$ has decidable equality, being zero and being a successor are both decidable. Then \cref{lem:cnf:limit} shows that $\cnf$ has classification. We then get classifiability induction from \cref{thm:CFI}.
\end{proof}

We use classifiability induction to prove the uniqueness of the arithmetic operations.
\begin{theorem}[\flinkFull{https://cj-xu.github.io/agda/type-theoretic-approaches-to-ordinals/index.html\#Theorem-28}]
\label{thm:cnf-arith-unique}
The operations of addition, multiplication and exponentiation with base $\omega$ on $\cnf$ are unique.
\end{theorem}
\begin{proof}
We sketch the proof for the uniqueness of addition and refer to our formalisation for the rest. Assume that $(+')$ is also an addition operation on CNFs. The goal is to show that $x+y = x+'y$ for all $x$ and $y$. We use classifiability induction on~$y$. The zero- and successor-cases are trivial. When $y$ is a limit, we use the fact that $(+)$ preserves suprema, i.e., if $a$ is a supremum of a sequence $f$, then $c+a$ is a supremum of the sequence $\lambda i. c+f_i$.
\end{proof}

We can check if a CNF is a limit and construct the fundamental sequence for limit CNFs. However, we cannot compute suprema or limits in general.

\begin{theorem}[\flinkFull{https://cj-xu.github.io/agda/type-theoretic-approaches-to-ordinals/index.html\#Theorem-29}]
  \label{thm:cnf-sups-lims-proof}
  $\cnf$ does not have suprema or limits.
  Assuming the law of exclude middle $(\LEM)$, $\cnf$ has suprema (and thus limits) of arbitrary \emph{bounded} sequences.
  If $\cnf$ has limits of bounded strictly increasing sequences, then the weak limited principle of omniscience $(\WLPO)$ is derivable.
\end{theorem}
\begin{proof}
To show that $\cnf$ does not have suprema or limits, we construct the following counterexample. Let $\omega \uparrow\uparrow$ be a sequence of CNFs defined by $\omega \uparrow\uparrow 0 \defeq \omega$ and $\omega \uparrow\uparrow (k+1) \defeq \omega^{\omega \uparrow\uparrow k}$. If it has a limit, say $x$, then any CNF $a$ is strictly smaller than $x$, including $x$ itself. But this is in contradiction with irreflexivity.

For the second part, we use Theorem~10.4.3 from the HoTT book~\cite[Thm~10.4.3]{hott-book} which we recall as Lemma~\ref{nonempty-subset-ord} in Section~\ref{sec:bookords}. It states that, assuming $\LEM$, $(A,\prec)$ is an extensional well-founded order if and only if every nonempty subset $B \subseteq A$ has a least element. Given a bounded sequence $f$ with bound $b$, we consider the subset $P : \cnf \to \Prop$ of all the CNFs that are upper bounds of $f$. This subset contains at least $b$ and is thus nonempty. We already have that $<$ on $\cnf$ is extensional and well-founded. Therefore, if we assume $\LEM$, then $P$ has a least element which is a supremum of $f$.

On the other hand, the proof of \cref{thm:no-go} demonstrates that, if sequences bounded by $\omega \cdot 2$ have limits, then $\WLPO$ holds.
\end{proof}

\section{Brouwer Trees}
\label{sec:HIIT}

We now consider the  construction of Brouwer trees in \cref{subsec:Brouwer-as-HIIT} in more detail: the type $\brouwer$ was defined mutually with the relation $\leq$, and we defined $x < y$ as $\bsuc \, x \leq y$.
The elimination principles for such a \emph{quotient inductive-inductive construction}~\cite{gabe:phd} are on an intuitive level explained in the HoTT book (e.g.\ in Chapter 11.3~\cite[Chp~11.3]{hott-book}), and a full specification as well as further explanations are given by Altenkirch, Capriotti, Dijkstra, Kraus and Nordvall Forsberg~\cite{Altenkirch2018} and Kaposi and Kov\'acs~\cite{kaposiKovacs:hiitSyntax,AAhiits,kaposi2019constructing}.

We want to elaborate on the arguments that are required to establish the results listed in \cref{sec:axiomaticapproach}.
Many proofs are very easy, for example the property (A\ref{item:assumption3}) of ``mixed transitivity'' is (almost) directly given by the constructor $\ltrans$ (cf.\ our formalisation);
the property \eqref{eq:leq-<-bad-direction} is true as well, with an only slightly less direct argument.
When we prove a propositional property by induction on a Brouwer tree, we only need to consider cases for point constructors, and multiple properties already follow from this.
Below, we focus on the more difficult arguments and explain some of the more involved proofs.

\subsection{Distinction of Constructors}
\label{subsec:distincton-of-constructors}

To start with, we need to prove that the point constructors of $\brouwer$ are distinguishable, e.g.\ that we have $\neg (\bzero = \bsuc \, x)$ --- point constructors are not always distinct in the presence of path constructors.
Nevertheless, this is fairly simple in our case, as the path constructor $\bbisim$ only equates limits, and the standard strategy of simply defining distinguishing families (such as $\isZero : \brouwer \to \Prop$ in the proof of \cref{lem:distincton-of-constructors} below) works.

\begin{lemma}[\flinkFull{https://cj-xu.github.io/agda/type-theoretic-approaches-to-ordinals/index.html\#Lemma-30}]
  \label{lem:distincton-of-constructors}
	The constructors of $\brouwer$ are distinguishable, i.e.\ one can construct proofs of $\bzero \neq \bsuc \, x$, $\bzero \neq \blimit \, f$, and $\bsuc \, x \neq \blimit \, f$.
\end{lemma}
\begin{proof}
We show how to distinguish $\bzero$ and $\bsuc\,x$; the other parts are shown in the same way.
Setting
\begin{alignat}{4}
&\isZero &\;& \bzero &\, & \defeq &\quad &  \Unit \\
&\isZero && (\bsuc \, x) && \defeq &&  \Empty \\
&\isZero && (\blimit \, f) && \defeq &&  \Empty 
\end{alignat}
means the proof obligations for the path constructors ($\bbisim$ and the truncation constructor) are trivial. Now if $\bzero = \bsuc \, x$, since $\isZero \; \bzero$ is inhabited, $\isZero\, (\bsuc \, x) \equiv \Empty$ must be as well --- a contradiction, which shows $\neg (\bzero = \bsuc \, x)$.
\end{proof}

\subsection{Codes Characterising \texorpdfstring{$\leq$}{<=}} \label{subsec:brw-codes}

Antisymmetry of $\leq$ as well as well-foundedness and extensionality of $<$ are among the technically most difficult results about $\brouwer$ that we present in this paper.
They are also the properties that would most easily fail with the ``wrong'' definition of $\brouwer$.
%
%
%
%
To see the difficulty, let us for example consider well-foundedness of $<$:
Given a strictly increasing sequence $f$, we have to show that $\blimit \, f$ is accessible, i.e.\ that any given $x < \blimit \, f$ is accessible.
However, the induction hypothesis only tells us that every $f \, k$ is accessible.
Thus, we want to show that there exists a $k$ such that $x < f \, k$, but doing this directly does not seem possible.

We use a strategy corresponding to the \emph{encode-decode method} \cite{licataShulman_circle} and define a type family
\begin{equation}
\Code : \brouwer \to \brouwer \to \Prop
\end{equation}
which has the \emph{correctness} properties
\begin{alignat}{5}
& \toCode : &\; & x \leq y \to \Code \, x \, y \\
& \fromCode : && \Code \, x \, y \to x \leq y,
\end{alignat}
for every $x, y : \brouwer$,
with the goal of providing a concrete description of $x \leq y$.
On the point constructors, the definition works as follows:
\newcommand{\noparen}{\phantom{(}}
\begin{alignat}{7}
&\Code &\;\;& \noparen\bzero &\;&  \noparen \blank &\; &\defeq &\quad &  \Unit \\
&\Code && (\bsuc \, x) &&  \noparen\bzero &&\defeq &&  \Empty \\
&\Code && (\bsuc \, x) &&  (\bsuc \, y) &&\defeq &&  \Code \; x \; y \label{eq:code-suc-suc}\\
&\Code && (\bsuc \, x) &&  (\blimit \, f) &&\defeq &&  \exists n. \Code \, (\bsuc \, x) \, (f \, n) \label{eq:code-suc-limit} \\
&\Code && (\blimit \, f) &&  \noparen\bzero &&\defeq &&  \Empty \\
&\Code && (\blimit \, f) &&  (\bsuc \, y) &&\defeq &&  \forall k. \Code \, (f \, k) \, (\bsuc \, y) \label{eq:code-limit-suc}\\
&\Code && (\blimit \, f) &&  (\blimit \, g) &&\defeq &&
\forall k. \exists n.  \Code \, (f \, k) \, (g \, n)
\label{eq:code-limit-limit}
\end{alignat}

The part of the definition of $\Code$ given above is easy enough; the tricky part is defining $\Code$ for the path constructor $\bbisim$.
If for example we have $g \approx h$, we need to show that $\Code \; (\blimit \, f) \, (\blimit \, g) \; = \; \Code \; (\blimit \, f) \, (\blimit \, h)$.
The core argument is not difficult; using the bisimulation $g \approx h$, one can translate between indices for $g$ and $h$ with the appropriate properties.
However, this example already shows why this becomes tricky: The bisimulation gives us inequalities $\leq$, but the translation requires instances of $\Code$, which means that $\toCode$ has to be defined \emph{mutually} with $\Code$.
This is still not sufficient: In total, the mutual higher inductive-inductive construction needs to simultaneously prove and construct $\Code$, $\toCode$, versions of transitivity and reflexivity of $\Code$ as well several auxiliary lemmas:
\begin{align}
  & \toCode : \; x \leq y \to \Code \, x \, y \\
  & \mathsf{Code\text{-}trans} : \Code\, x\, y \to \Code\, y\, z \to \Code\, x\, z \\
  & \mathsf{Code\text{-}refl} : \Code\, x\, x \\
  & \mathsf{Code\text{-}cocone} : \Code\, x\,(f k) \to \Code\, x\,(\blimit\, f) \\
  &  \mathsf{Code\text{-}succ\text{-}incr\text{-}simple} : \Code\, x\, (\bsuc\, x)
\end{align}
After the mutual definition is complete, we can separately prove $\fromCode : \Code \, x \, y \to x \leq y$.
The complete construction is presented in the Agda formalisation.


$\Code$ allows us to easily derive various useful auxiliary lemmas, for example the following four:

\begin{lemma}[\flinkFull{https://cj-xu.github.io/agda/type-theoretic-approaches-to-ordinals/index.html\#Lemma-31}]
  \label{lem:suc-mono-inv}
    For $x,y : \brouwer$, we have $x \leq y \leftrightarrow \bsuc \, x \leq \bsuc \, y$.
\end{lemma}
\begin{proof}
    This is a direct consequence of \eqref{eq:code-suc-suc} and the correctness of $\Code$.
\end{proof}

\begin{lemma}[\flinkFull{https://cj-xu.github.io/agda/type-theoretic-approaches-to-ordinals/index.html\#Lemma-32}]
  \label{lem:x-under-limit-means-x-under-element}
    Let $f : \N \toincr{<} \brouwer$ be a strictly increasing sequence and $x : \brouwer$ a Brouwer tree such that $x < \blimit \, f$.
    Then, there exists an $n : \N$ such that $x < f \, n$.
\end{lemma}
\begin{proof}
    By definition, $x < \blimit \, f$ means $\bsuc \, x \leq \blimit \, f$.
    Using $\toCode$
    together with the case
    \eqref{eq:code-suc-limit},
    there exists an $n$ such that $\Code\,(\bsuc\,x)\, (f \, n)$.
    Using $\fromCode$, we get the result.
\end{proof}

\begin{lemma}[\flinkFull{https://cj-xu.github.io/agda/type-theoretic-approaches-to-ordinals/index.html\#Lemma-33}]
  \label{lem:lim-under-lim-simulation}
    If $f$, $g$ are strictly increasing sequences with $\blimit \, f \leq \blimit \, g$, then $f$ is simulated by $g$.
\end{lemma}
\begin{proof}
    For every $k : \N$, \eqref{eq:code-limit-limit} tells us that there exists an $n : \N$ such that, after using $\fromCode$, we have $f \, k \leq g \, n$.
\end{proof}

\begin{lemma}[\flinkFull{https://cj-xu.github.io/agda/type-theoretic-approaches-to-ordinals/index.html\#Lemma-34}]
  \label{lem:lim-under-suc}
    If $f$ is a strictly increasing sequence and $x$ a Brouwer tree such that $\blimit \, f \leq \bsuc \, x$, then $\blimit \, f \leq x$. Dually, limits are closed under successors: if $x < \blimit\,f$ then also $\bsuc\,x < \blimit\,f$.
\end{lemma}
\begin{proof}
  From \eqref{eq:code-limit-suc}, we have that $f\,k \leq \bsuc \, x$ for every $k$. But since $f$ is increasing, $\bsuc\,(f\,k) \leq f\,(k + 1) \leq \bsuc \, x$ for every $k$, hence by \cref{lem:suc-mono-inv} $f\,k \leq x$ for every $k$, and the result follows using the constructor $\llimiting$.
  For the second statement, if $x < \blimit\,f$ then by \cref{lem:x-under-limit-means-x-under-element} we have $x < f\,n$ for some $n$, and since $f$ is increasing, $\bsuc\,x < f\,(n + 1) < \blimit\,f$.
\end{proof}

An alternative proof of \cref{lem:suc-mono-inv}, which does not rely on the machinery of codes, is given in our formalisation.

\subsection{Antisymmetry, Well-Foundedness, and Extensionality}

With the help of 
the consequences of the characterisation of $\leq$ shown above,
we can show multiple non-trivial properties of $\brouwer$ and its relations.
Regarding well-foundedness, we can now complete the argument sketched above:

\begin{theorem}[\flinkFull{https://cj-xu.github.io/agda/type-theoretic-approaches-to-ordinals/index.html\#Theorem-35}]
  \label{thm:brouwer-wellfounded}
    The relation $<$ is well-founded.
\end{theorem}
\begin{proof}
    We need to prove that every $y : \brouwer$ is accessible.
    When doing induction on $y$, the cases of path constructors are automatic as we are proving a proposition.
    From the remaining constructors, we only show the hardest case, which is when $y \equiv \blimit \, f$.
    We have to prove that any given $x < \blimit \, f$ is accessible.
    By \cref{lem:x-under-limit-means-x-under-element}, there exists an $n$ such that $x < f\, n$, and the latter is accessible by the induction hypothesis.
\end{proof}

Next we show that $\leq$ is antisymmetric, i.e.\ if $x \leq y$ and $y \leq x$ then $x = y$.

\begin{theorem}[\flinkFull{https://cj-xu.github.io/agda/type-theoretic-approaches-to-ordinals/index.html\#Theorem-36}]
  \label{thm:brouwer-antisymmetric}
    The relation $\leq$ is antisymmetric.
\end{theorem}
\begin{proof}
    Let $x, y$ with $x \leq y$ and $y \leq x$ be given.
    We do nested induction. As before, we can disregard the cases for path constructors, giving us $9$ cases in total, many of which are duplicates.
    We discuss the two most interesting cases:
    \begin{itemize}
        \item $x \equiv \blimit \, f$ and $y \equiv \bsuc \, y'$: In that case, \cref{lem:lim-under-suc} and the assumed inequalities show $\bsuc \, y' \leq \blimit \,f \leq y'$ and thus $y' < y'$, contradicting the well-foundedness of $<$.
        \item $x \equiv \blimit \, f$ and $y \equiv \blimit \, g$: By \cref{lem:lim-under-lim-simulation}, $f$ and $g$ simulate each other. By the constructor $\bbisim$, $x = y$. \qedhere
    \end{itemize}
\end{proof}

\begin{corollary}[\flinkFull{https://cj-xu.github.io/agda/type-theoretic-approaches-to-ordinals/index.html\#Corollary-37}]
  \label{cor:brouwer-satisfies-assumptions}
	$(\brouwer, <, \leq)$ satisfies the assumptions (A\ref{item:assumption1}), (A\ref{item:assumption2}), and (A\ref{item:assumption3}), i.e.\ it is an instantiation of the abstract triple $(A,<,\leq)$ discussed in \cref{sec:axiomaticapproach}. Furthermore, the symmetric variation of the second half of assumption (A\ref{item:assumption3}) holds for $(\brouwer, <, \leq)$, i.e.,\ if $x \leq y$ and $y < z$, then $x < z$.
\end{corollary}
\begin{proof}
	The only non-immediate properties are antisymmetry of $\leq$ (\cref{thm:brouwer-antisymmetric}) and irreflexivity of $<$ (\cref{thm:brouwer-wellfounded} and \cref{lem:no-infinite-sequence}). That $x \leq y < z$ implies $x < z$ follows directly from the definition of $a < b$ as $\bsuc\,a \leq b$, transitivity of $\leq$, and monotonicity of $\bsuc$.
\end{proof}

Finally we can show that $<$ is extensional, i.e.\ that Brouwer trees
with the same predecessors are equal.

\begin{theorem}[\flinkFull{https://cj-xu.github.io/agda/type-theoretic-approaches-to-ordinals/index.html\#Theorem-38}]
  \label{thm:brouwer-extensional}
    The relation $<$ is extensional.
\end{theorem}
\begin{proof}
    Let $x$ and $y$ be two elements of $\brouwer$ with the same set of smaller elements.
    As in the above proof, 
    we can consider $9$ cases. If $x$ and $y$ are built of different constructors, it is easy to derive a contradiction. For example, in the case $x \equiv \blimit \, f$ and $y \equiv \bsuc \, y'$, we have $y' < y$ and thus $y' < \blimit \, f$.
    By \cref{lem:x-under-limit-means-x-under-element},
    there exists an $n$ such that $y' < f \, n$, which in turn implies $y < f \, (n+1)$ and thus $y < \blimit \, f$. By the assumed set of smaller elements, that means we have $y < y$, contradicting well-foundedness.
    
    The other interesting case, $x \equiv \blimit \, f$ and $y \equiv \blimit \, g$, is easy. For all $k : \N$, we have $f \, k < f \, (k+1) \leq \blimit \, f$ and thus $f \, k < \blimit \, g$; by the constructor $\llimiting$, this implies $\blimit \, f \leq \blimit \, g$.
    By the symmetric argument and by antisymmetry of $\leq$, it follows that $\blimit \, f = \blimit \, g$.
\end{proof}

\subsection{Classifiability}

Classifiability is straightforward for $\brouwer$, as the point constructors of the data type exactly corresponds to zero, successors and limits.

\begin{lemma}[\flinkFull{https://cj-xu.github.io/agda/type-theoretic-approaches-to-ordinals/index.html\#Lemma-39}]
  \label{thm:brouwer-has-it-all}
    $\brouwer$ has zero, strong successors, and limits of strictly increasing sequences, and each part is given by the corresponding constructor.\qed
\end{lemma}
\begin{proof}
  Most of these claims are easy.
  To verify that $\bsuc$ is a strong successor we need to show that $x < \bsuc \, b$ implies $x \leq b$.
  But $x < \bsuc \, b$ is defined to mean $\bsuc \, x \leq \bsuc \, b$ which, by \cref{lem:suc-mono-inv}, is indeed equivalent to $x \leq b$.
\end{proof}

By the definition of $<$ for $\brouwer$, we have:
\begin{corollary}[\flinkFull{https://cj-xu.github.io/agda/type-theoretic-approaches-to-ordinals/index.html\#Corollary-40}, of \cref{lem:suc-mono-inv}] \label{cor:brw-suc-mono}
	The strong successor of $\brouwer$ is $<$- and $\leq$-monotone.
\end{corollary}

Hence we can now observe that the special case of induction for $\brouwer$ where the goal is a proposition is exactly classifiability induction, and by \cref{thm:CFI-to-classification}, $\brouwer$ has classification. This proves the following theorem:

\begin{theorem}[\flinkFull{https://cj-xu.github.io/agda/type-theoretic-approaches-to-ordinals/index.html\#Theorem-41}]
  \label{thm:brw-class-CFI}
    $\brouwer$ has classification and satisfies classifiability induction. \qed
\end{theorem}

\subsection{Arithmetic of Brouwer Trees}

The standard arithmetic operations on Brouwer trees can be implemented with the usual strategy well-known in the functional programming community, i.e.\ by recursion on the second argument.
However, there are several additional difficulties which stem from the fact that our Brouwer trees enforce correctness.

Let us start with addition.
The obvious definition is
\begin{alignat}{10}
&x &\; & + &\; & \bzero  &\quad & \defeq  &\quad & x \label{eq:brw-adding-zero}\\
&x && + && \bsuc \, y  && \defeq  && \bsuc \, (x + y) \\
&x && + && \blimit \, f  && \defeq  && \blimit \, (\lambda k. x + f \, k).
\end{alignat}
For this to work, we need to prove, mutually with the above definition, that the sequence $\lambda k. x + f\, k$ in the last line is still increasing, which follows from mutually proving that $+$ is monotone in the second argument, both with respect to $\leq$ and $<$.
We also need to show that bisimilar sequences $f$ and $g$ lead to bisimilar sequences $x + f\, k$ and $x + g\, k$.

The same difficulties occur for multiplication $(\cdot)$, where they are more serious:
Even if $f$ is increasing (with respect to $<$), then
$\lambda k.\, x \cdot f \, k$ is not necessarily increasing, as $x$ could be $\bzero$.
What saves us is that it is \emph{decidable} whether $x$ is $\bzero$ (cf.~\cref{subsec:distincton-of-constructors}); and if it is, the correct definition is $x \cdot \blimit \, f \defeq \bzero$.
If $x$ is not $\bzero$, then it is at least $\bsuc \, \bzero$ (another simple lemma for $\brouwer$), and the sequence \emph{is} increasing.
With the help of several lemmas that are all stated and proven mutually with the actual definition of $(\cdot)$, 
the mentioned decidability is the core ingredient which allows us to complete the construction:
\begin{alignat}{10}
&x &\; & \cdot &\; & \bzero  &\quad & \defeq  &\quad & \bzero \\
&x && \cdot && \bsuc \, y  && \defeq  && (x \cdot y) + x \\
&x && \cdot && \blimit \, f  && \defeq  &&
                                           \begin{cases}
                                             \bzero & \text{if $x = \bzero$} \\
                                             \blimit \, (\lambda k. x \cdot f \, k). & \text{otherwise} \\
                                           \end{cases}
\end{alignat}
That $\lambda k. x \cdot f \, k$ is increasing if $x > \bzero$ and $f$ is increasing follows from mutually proving that $\cdot$ is monotone in the second argument, and that $\bzero \cdot y = \bzero$.
Exponentiation $x^y$ comes with similar caveats as multiplication, but works with the same strategy.
\begin{alignat}{10}
&x^ \bzero  &\quad & \defeq  &\quad & \bsuc\,\bzero \\
&x^ {\bsuc \, y}  && \defeq  && (x^y) \cdot x \\
&x^{\blimit \, f}  && \defeq  &&
                                           \begin{cases}
                                             \bzero & \text{if $x = \bzero$} \\
                                             \bsuc\,\bzero & \text{if $x = \bsuc\, \bzero$} \\
                                             \blimit \, (\lambda k. x^{f \, k}) & \text{otherwise.}
                                           \end{cases} \label{eq:brw-exp-limit}
\end{alignat}

With these definitions, the properties introduced in \cref{subsec:abstract-arithmetic} are automatically satisfied, and our Agda formalisation shows that these properties describe the above operations uniquely:
\begin{theorem}[\flinkFull{https://cj-xu.github.io/agda/type-theoretic-approaches-to-ordinals/index.html\#Theorem-42}]
  \label{thm:brw-has-all-arithmetic-properties}
	$\brouwer$ has unique addition, multiplication, and exponentation, given by \eqref{eq:brw-adding-zero} -- \eqref{eq:brw-exp-limit}. \qed
\end{theorem}

Many arithmetic properties can easily be established by induction, for example:
\begin{lemma}[\flinkFull{https://cj-xu.github.io/agda/type-theoretic-approaches-to-ordinals/index.html\#Lemma-43-i}]
  \label{lem:brw-arith-properties}
  \mbox{}
  \begin{enumerate}[(i)]
  \item Addition and multiplication are weakly monotone in the first argument: if $x \leq y$ then $x + z \leq y + z$ and $x \cdot z \leq y \cdot z$.
  \item Addition is left cancellative: if $x + y = x + z$ then $y = z$.
  \item Addition and multiplication associative, and multiplication distributes over addition: $x \cdot (y + z) = (x \cdot y) + (x \cdot z)$.
  \item Exponentiation is a homomorphism: $x^{y + z} = x^y \cdot x^z$. \qed
  \end{enumerate}
\end{lemma}

The first infinite ordinal $\omega$ can be defined as a Brouwer tree as $\omega \defeq \blimit\, \iota$, where $\iota : \Nat \toincr{<} \brouwer$ embeds the natural numbers as finite Brouwer trees. As an example of how $\brouwer$ behaves as expected as a type of ordinals, we can also define $\varepsilon_0 \defeq \blimit\,(\lambda k.\omega \uparrow\uparrow k)$, where $\omega \uparrow\uparrow (k+1) \defeq \omega^{\omega \uparrow\uparrow k}$, and use the $\bbisim$ constructor to show that indeed $\omega^{\varepsilon_0} = \varepsilon_0$. Similarly, using the lemmas we have already established, it is now not so hard to establish other expected properties of Brouwer trees, that would not hold without the path constructors in the definition of $\brouwer$:

\begin{lemma}[\flinkFull{https://cj-xu.github.io/agda/type-theoretic-approaches-to-ordinals/index.html\#Lemma-44}]
    \label{thm:additive-principal}
    Brouwer trees of the form $\omega^x$ are additive principal:
    \begin{enumerate}[(i)]
        \item if $a < \omega^x$ then $a + \omega^x = \omega^x$. \label{item:add-prin-1}
        \item if $a < \omega^x$ and $b < \omega^x$ then $a + b < \omega^x$. \label{item:add-prin-2}
    \end{enumerate}
        Furthermore, if $x > \bzero$ and $n : \Nat$, then $\iota(n+1) \cdot \omega^x = \omega^x$.     \label{thm:omega-preserves-finite}
\end{lemma}
\begin{proof}
    For \eqref{item:add-prin-1}, by antisymmetry it is enough to show $a + \omega^x \leq \omega^x$, since $\omega^x \leq a + \omega^x$ holds by $\omega^x = 0 + \omega^x$ and monotonicity of addition. We prove $a + \omega^x \leq \omega^x$ by induction on $a$, making crucial use of 
    \cref{lem:x-under-limit-means-x-under-element}.
    Statement \eqref{item:add-prin-2} is an easy corollary of \eqref{item:add-prin-1} and strict monotonicity of $+$ in the second argument.
    The final statement is proven by induction on $x$, with an inner induction on $n$ for the successor case.
\end{proof}

Perhaps more surprisingly, even though $\brouwer$ has addition with expected properties, it is a constructive taboo that $\brouwer$ has subtraction. This is in contrast with the situation for $\cnf$, where subtraction is computable, and in fact was crucial in our proof of correctness of the arithmetic operations.

\begin{theorem}[\flinkFull{https://cj-xu.github.io/agda/type-theoretic-approaches-to-ordinals/index.html\#Theorem-45}]
  \label{thm:brw-subtraction-iff-lpo}
  If $\brouwer$ has subtraction, then it has unique subtraction.
  $\brouwer$ has subtraction if and only if $\LPO$ holds.
\end{theorem}
\begin{proof}
Firstly, note that the type of having
  subtraction for $\brouwer$ is a proposition due to left cancellability of addition:
  if $z$ and $z'$ are candidates for $y - x$, then
  $x + z = y = x + z'$, hence $z = z'$ by
  \cref{lem:brw-arith-properties}(ii). Hence subtraction and unique subtraction coincide for $\brouwer$.

  In \cref{thm:brw-lpo-iff-trich-and-split}, we will show that $\LPO$ holds if and
  only if $\Splits(\brouwer, <, \leq)$. Thus it is sufficient to show
  that $\brouwer$ has subtraction if and only if
  $\Splits(\brouwer, <, \leq)$. If $\brouwer$ has subtraction, then we
  can split $p : x \leq y$ by comparing $y -_{p} x$ with $0$, which is
  possible by \cref{lem:distincton-of-constructors} --- we have
  $x = y$ if $y -_{p} x = 0$, and $x < y$ if $y -_{p} x >
  0$. Conversely, assume $\leq$ splits, and let $x$, $y$ with
  $x \leq y$ be given. Since having subtraction for $\brouwer$ is a proposition, we can use classifiability induction on $y$ to
  construct $y -_{p} x$. In each case, we first split $p$: if $x = y$,
  then we define $y -_{p} x \defeq \bzero$. If instead $x < y$, we
  cannot have $y = \bzero$, and if $y = \bsuc\,y'$, we can use
  \cref{lem:suc-mono-inv} to define $y -_{p} x$ using the
  induction hypotheses. Finally if $y = \blimit\,f$ we use
  \cref{lem:x-under-limit-means-x-under-element} to again be able to
  use the induction hypothesis to finish the job.
\end{proof}

\subsection{Decidability and Undecidability for Brouwer Trees}
\label{subsec:Brw-decidability}


We now consider what is decidable and what is not for Brouwer
trees. Because we can distinguish constructors by
\cref{lem:distincton-of-constructors}, we can decide most properties
of finite Brouwer trees, i.e.\ Brouwer trees $x$ with $x < \omega$.

\begin{theorem}[\flinkFull{https://cj-xu.github.io/agda/type-theoretic-approaches-to-ordinals/index.html\#Theorem-46}]
  \label{thm:brw-decidable}
	It is decidable whether a Brouwer tree is finite. If $n$ is a finite Brouwer tree and $\sim$ is one of the relations $=$, $<$, $\leq$, then the predicates $(n \sim \blank)$ and $(\blank \sim n)$ are decidable.
\end{theorem}
\begin{proof}
  Deciding finiteness is easy to do by induction, since $\blimit\,f$
  is never finite, and hence the $\bbisim$ constructor is trivially
  respected. Similar considerations apply when deciding
  $(n \sim \blank)$ and $(\blank \sim n)$; see the Agda formalisation for details.
\end{proof}

\cref{thm:brw-decidable} covers the most important properties that are decidable.
In order to demonstrate that certain other properties cannot be shown to be decidable, we use constructive taboos as discussed in \cref{subsec:taboos}.
Many constructive taboos talk about binary sequences, while the sequences that are important in the construction of $\brouwer$ are strictly increasing sequences of Brouwer trees.
In order connect the taboos with properties of $\brouwer$, we therefore want to be able to translate between both types of sequences.
Recall the construction of the \emph{jumping sequence} in the proof of \cref{thm:no-go}.
For $\brouwer$, we can implement it as a concrete function of type
\begin{equation} \label{eq:jumping-sequence}
\jump{\cdot} \colon (\Nat \to \Bool) \to (\N \toincr{<} \brouwer).
\end{equation}
We then have:
\begin{lemma}[\flinkFull{https://cj-xu.github.io/agda/type-theoretic-approaches-to-ordinals/index.html\#Lemma-47}]
  \label{lem:jumping-omega-vs-true}
	For any binary sequence $s : \N \to \Bool$, we have
	$\blimit \, \jump s \leq \omega \cdot 2$. Moreover,
	the following three statements are equivalent:
	\begin{enumerate}[(i)]
	\item $\exists k. s_k = \btt$ \label{item:lem-jump-1} 
	\item $\blimit \, \jump s = \omega \cdot 2$ \label{item:lem-jump-2} 
	\item $\omega < \blimit \, \jump s$ \label{item:lem-jump-3} 
\end{enumerate}
\end{lemma}
\begin{proof}
For any $i$, we have $\jump s_i \leq \omega + i$, and thus 	$\blimit \, \jump s \leq \omega \cdot 2$.
To see the equivalences, we have:

$\eqref{item:lem-jump-1}\Rightarrow\eqref{item:lem-jump-2}$:
From \eqref{item:lem-jump-1}, we can compute a minimal $k$
such that $s_k = \btt$. Then, we have $\jump s_i = i$ for $i < k$, and $\jump s_{k+i} = \omega + i$.
Therefore, $\blimit \, \jump s = \omega \cdot 2$.

$\eqref{item:lem-jump-2}\Rightarrow\eqref{item:lem-jump-3}$:
This is immediate as $\omega < \omega \cdot 2$.

$\eqref{item:lem-jump-3}\Rightarrow\eqref{item:lem-jump-1}$:
By \cref{lem:x-under-limit-means-x-under-element}, there exists some $n$ such that $\omega < \jump s \, n$.
In particular, $\jump s \, n$ is infinite.
By \cref{lemma:sigma-lpo-vs-lpo} and \cref{thm:brw-decidable}, we can therefore find an $n$ such that $\jump s \, n$ is infinite. The minimal such $n$ has the property that $s_n = \btt$.
\end{proof}	

Vice versa, assume $f \colon \Nat \to \brouwer$ is a sequence and $P : \brouwer \to \Prop$ a predicate such that for all $n$, $P(f \, n)$ is decidable.
We then define the \emph{unjumping sequence}
\begin{equation}
\unjump f P \colon \Nat \to \Bool
\end{equation}
by setting
\begin{equation} \label{eq:inv-jumping-sequence-construction}
\unjump f P \, n \defeq
\begin{cases}
\btt & \textit{if $P(f \, n)$} \\
\bff & \textit{if $\neg P(f \, n)$.}
\end{cases}
\end{equation}

We now put the jumping sequence to work to show that most decidability questions of arbitrary Brouwer trees are in fact equivalent to each other, and to the constructive taboo of \LPO.

\begin{theorem}[\flinkFull{https://cj-xu.github.io/agda/type-theoretic-approaches-to-ordinals/index.html\#Theorem-48}]
  \label{thm:brw-LPO}
	For the type of Brouwer trees, the following statements are equivalent:
	\begin{enumerate}[(i)]
		\item $\LPO$ \label{item:brw-lpo} 
		\item $\forall x,y. \Dec(x \leq y)$ \label{item:brw-leq-dec}
		\item $\forall x,y. \Dec(x < y)$ \label{item:brw-<-dec}
		\item $\forall x. \Dec(\omega < x)$ \label{item:brw-w<-dec}
		\item $\forall x,y. \Dec (x = y)$ \label{item:brw-eq-dec}
		\item $\forall x. \Dec (x = \omega \cdot 2)$ \label{item:brw-=w+2-dec}
	\end{enumerate}
\end{theorem}
\begin{proof}
	We show the equivalence using two cycles:
	$\eqref{item:brw-lpo} \Rightarrow \eqref{item:brw-leq-dec} \Rightarrow \eqref{item:brw-<-dec} \Rightarrow \eqref{item:brw-w<-dec} \Rightarrow \eqref{item:brw-lpo}$
	and
	$\eqref{item:brw-lpo} \Rightarrow \eqref{item:brw-eq-dec} \Rightarrow \eqref{item:brw-=w+2-dec} \Rightarrow \eqref{item:brw-lpo}$.
	
	$\eqref{item:brw-lpo} \Rightarrow \eqref{item:brw-leq-dec}$: 
	Assume $\LPO$ and define $P(x) \defeq \forall y. \Dec(x \leq y)$. We show $\forall x. P(x)$ by classifiability induction.
	
	\begin{itemize}
		\item case $P(\bzero)$: Trivial, since $\bzero \leq y$ for any $y$.
		\item case $P(\bsuc \, x)$: To decide $\bsuc \, x \leq y$, we do classifiability induction on $y$.
		Using the results of \cref{subsec:brw-codes}, we know that $\bsuc \, x \leq \bzero$ is false and that $\bsuc \, x \leq \bsuc \, y'$ holds iff $x \leq y'$, which is decidable by the induction hypothesis.
		Asked whether $\bsuc \, x \leq \blimit \, f$, we use that $\bsuc \, x \leq f \, n$ is decidable by the induction hypothesis and consider the unjumping sequence $\unjump{f}{Q}$, with $Q(z) \defeq (\bsuc \, x \leq z)$; i.e.\ we have
		\begin{equation}
		\unjump{f}{Q}(n) \defeq
		\begin{cases}
		  \btt & \textit{if $\bsuc \, x \leq f\, n$} \\
		  \bff & \textit{if $\neg(\bsuc \, x \leq f\, n)$.}
		\end{cases}
		\end{equation}
		Thanks to $\LPO$, we know that the sequence $\unjump{f}{Q}$ is either constantly $\bff$, or (cf.~\cref{lemma:sigma-lpo-vs-lpo}) we get $n$ such that $\bsuc \, x \leq f\, n$.
		In the first case, we assume $\bsuc \,x \leq \blimit \, f$; by \cref{lem:x-under-limit-means-x-under-element}, this implies $\exists n. \bsuc \, x \leq f\, n$, contradicting the statement that the sequence is constantly $\bff$.
		In the second case, we clearly have $\bsuc \,x \leq \blimit \, f$.
		\item case $P(\blimit \, f)$:
		We have to decide $\blimit \, f \leq y$. We use the unjumping sequence with $Q(z) \defeq \neg(z \leq y)$, i.e.\
		\begin{equation}
		\unjump{f}{Q}(n) \defeq
		\begin{cases}
		\btt & \textit{if $\neg(f\, n \leq y)$} \\
		\bff & \textit{if $f\, n \leq y$}
		\end{cases}
		\end{equation}
		Again, we apply $\LPO$. If the sequence is constantly $\bff$, we have $\blimit \, f \leq y$. If we get an $n$ with $\neg(f \, n \leq y)$, then the assumption $\blimit \, f \leq y$ gives a contradiction.
	\end{itemize}

	$\eqref{item:brw-leq-dec}\Rightarrow\eqref{item:brw-<-dec}$: Trivial, since $(x < y) \equiv (\bsuc \, x \leq y)$.

	$\eqref{item:brw-<-dec}\Rightarrow\eqref{item:brw-w<-dec}$: The latter is a special case of the former.
	
	$\eqref{item:brw-w<-dec}\Rightarrow\eqref{item:brw-lpo}$: Assume \eqref{item:brw-w<-dec} and let $s$ be a binary sequence. By assumption, we can decide $\omega < \blimit \, \jump s$, and thus $\exists i. s_i = \btt$ by \cref{lem:jumping-omega-vs-true}, implying $\LPO$.
	
	$\eqref{item:brw-lpo}\Rightarrow\eqref{item:brw-eq-dec}$: Since $(x = y) \leftrightarrow (x \leq y) \wedge (y \leq x)$ by antisymmetry, this follows from the result $\eqref{item:brw-lpo} \Rightarrow \eqref{item:brw-leq-dec}$ above.
	
	$\eqref{item:brw-eq-dec}\Rightarrow\eqref{item:brw-=w+2-dec}$: The latter is a special case of the former.
	
	$\eqref{item:brw-=w+2-dec}\Rightarrow\eqref{item:brw-lpo}$:
	Similar to the direction $\eqref{item:brw-w<-dec}\Rightarrow\eqref{item:brw-lpo}$ above, this follows from \cref{lem:jumping-omega-vs-true}.
\end{proof}

It is worth noting that the argument of $\eqref{item:brw-w<-dec}\Rightarrow\eqref{item:brw-lpo}$ in the above proof shows $\LPO$, while \cref{thm:no-go}, under similar assumptions and with a similar strategy, only shows $\WLPO$.
The construction of a concrete $n$ is made possible by the earlier results on $\brouwer$, but is not possible for the abstract situation considered by \cref{thm:no-go}.

As we have just seen, being able to check equality with $\omega \cdot 2$ is equivalent to $\LPO$, and equality with a finite number is always decidable.
Equality with $\omega$ lies in-between, in the sense that it is equivalent to $\WLPO$:

\begin{theorem}[\flinkFull{https://cj-xu.github.io/agda/type-theoretic-approaches-to-ordinals/index.html\#Theorem-49}]
  \label{thm:brw-dec-equ}
	$\brouwer$ has locally decidable equality at $\omega$ if and only if $\WLPO$ holds:
	\begin{equation}
	\WLPO \; \leftrightarrow \; \forall (x: \brouwer).\Dec(x = \omega)
	\end{equation}
\end{theorem}
\begin{proof}
	Assume $\forall x. \Dec(x = \omega)$. Let $s$ be a binary sequence. If $\blimit \, \jump s = \omega$, then $s$ is constantly $\bff$; otherwise, $s$ is not constantly $\bff$.
		
	Assume now $\WLPO$ and let $x: \brouwer$ be given. If $x$ is $\bzero$ or a successor, then $x \not= \omega$; thus, we assume that $x$ is $\blimit \, f$. Consider $\unjump{f}{\neg\mathsf{isFinite}}$.
	If this sequence is constantly $\bff$, then every $f_i$ is finite and the limit is $\omega$.
	If is it not constantly $\bff$, then $x$ must differ from $\omega$.
\end{proof}

For comparison, we observe the following:

\begin{theorem}[\flinkFull{https://cj-xu.github.io/agda/type-theoretic-approaches-to-ordinals/index.html\#Theorem-50}]
  \label{thm:brw-dec-equ-omega-times-n}
	Let $n$ be a natural number larger or equal to $2$.
	Deciding equalities locally at $\omega \cdot n$ is equivalent to $\LPO$:
	\begin{equation}
	\LPO \; \leftrightarrow \; \forall (x: \brouwer).\Dec(x = \omega \cdot n).
	\end{equation}	
\end{theorem}
\begin{proof}
	By left cancellation of addition (\cref{lem:brw-arith-properties}), we have
	\begin{equation}
	\left(\forall x. \Dec(x = \omega + a)\right) \to \left(\forall x. \Dec(x = a)\right)
	\end{equation}
	for any $a$.
	As a consequence, decidability of equality with $\omega \cdot n$, for $n \geq 2$, implies decidability of equality with $\omega \cdot 2$, which implies $\LPO$, which implies decidability of equality by \cref{thm:brw-LPO}.
\end{proof}

Summarising \cref{thm:brw-decidable,thm:brw-dec-equ,thm:brw-dec-equ-omega-times-n}, we have shown that decidability of equality with $\omega \cdot n$ holds for $n=0$, corresponds to $\WLPO$ for $n=1$, and to $\LPO$ if $n \geq 2$.
For stability, which is  somewhat weaker than decidability, we have:

\begin{theorem}[\flinkFull{https://cj-xu.github.io/agda/type-theoretic-approaches-to-ordinals/index.html\#Theorem-51}]
  \label{thm:brw-stable-equ}
	Equality of $\brouwer$ is stable at $\omega$, but stability at $\omega \cdot n$ for $n \geq 2$ implies $\MP$:
	\begin{align}
	& \forall (x: \brouwer).\Stable(x = \omega) \\
	& \left(\forall (x: \brouwer).\Stable(x = \omega \cdot n)\right) \to \MP.
	\end{align}
\end{theorem}
\begin{proof}
		Assume $\neg\neg(x = \omega)$; then, since $x$ cannot be $\bzero$ or a successor, let $x = \blimit \, f$.
		We have $x = \omega$ if and only if every $f_i$ is finite.
		If any $f_i$ is not finite, then $x \not= \omega$, in contradiction to the assumption.
		
		For the second part, setting $x \defeq \blimit \, \jump s$ and applying \cref{lem:jumping-omega-vs-true} shows immediately that the statement $\forall (x: \brouwer).\Stable(x = \omega \cdot 2)$ implies $\MP$. The general case for $n > 2$ again follows by left cancellation of addition (\cref{lem:brw-arith-properties}), since it implies
	\begin{equation}
	\left(\forall x. \Stable(x = \omega + a)\right) \to \left(\forall x. \Stable(x = a)\right)
	\end{equation}
	for any $a$.
\end{proof}
Adding a finite number $k$ does not change the situation in \cref{thm:brw-dec-equ,thm:brw-dec-equ-omega-times-n,thm:brw-stable-equ}:
If we replace $\omega \cdot n$ by $\omega \cdot n + k$, the remaining statements hold without any further difference.

The following observation will be the key ingredient of a proof that $\LPO$ implies trichotomy:
\begin{lemma}[\flinkFull{https://cj-xu.github.io/agda/type-theoretic-approaches-to-ordinals/index.html\#Lemma-52}]
  \label{lem:lpo-to-not-leq-lt}
	If $\LPO$ holds then, for all $x,y : \brouwer$, we have
	$\neg(x \leq y) \to (y < x)$.
\end{lemma}	
\begin{proof}
  Assume $\LPO$. Note that by \cref{lem:lpo-equiv-lpo-mp}, we then also have $\MP$. We do classifiability induction on $x$.
	\begin{itemize}
		\item Case $x \equiv \bzero$: The assumption $\neg(\bzero \leq y)$ is absurd.
		\item Case $x \equiv \bsuc \, x'$: We have to show $\neg(\bsuc \, x' \leq y) \to (y < \bsuc \, x')$.
		We do classifiability induction on $y$. The case $y \equiv \bzero$ is trivial, and the case $y \equiv \bsuc \, y'$ follows from \cref{lem:suc-mono-inv}.
		Finally, we need to consider the situation $y \equiv \blimit \, g$:
		\begin{equation}
		\begin{alignedat}{4}
		& \neg(\bsuc \, x' \leq \blimit \, g) & \quad & \Rightarrow & \quad & \neg \exists i. \bsuc \, x' \leq g_i \\
		&&& \Rightarrow && \forall i. \neg(\bsuc \, x' \leq g_i)\\ 
		&&& \Rightarrow && \forall i. g_i < \bsuc \, x' \\ 
		&&& \Rightarrow && \forall i. g_i \leq x' \\ 
		&&& \Rightarrow && \blimit \, g \leq x' \\ 
		&&& \Rightarrow && \blimit \, g < \bsuc \, x'.
		\end{alignedat}
		\end{equation}
		\item The final case is $x \equiv \blimit \, f$:
		\begin{equation}
		\begin{alignedat}{4}
		& \neg(\blimit \, f \leq y) & \quad & \Rightarrow & \quad & \neg(\forall i. f_i \leq y) \\
		&&& \Rightarrow && \neg\neg(\exists i. \neg(f_i \leq y))\\ 
		&&& \Rightarrow && \neg\neg(\exists i. y < f_i)\\ 
		&&&  && \textit{(using $\MP$ on a decidable family of propositions)}\\ 
		&&& \Rightarrow && \exists i. y < f_i\\ 
		&&& \Rightarrow && y < \blimit \, f.
		\end{alignedat}
		\end{equation}\qedhere
	\end{itemize}
\end{proof}

Using this lemma, and going via $\LPO$, we can now show that splitting $\leq$ implies trichotomy for $\brouwer$ --- in the general setting, only the reverse direction, i.e., that trichotomy implies splitting, holds.

\begin{theorem}[\flinkFull{https://cj-xu.github.io/agda/type-theoretic-approaches-to-ordinals/index.html\#Theorem-53}]
  \label{thm:brw-lpo-iff-trich-and-split}
	The following properties are equivalent for the type of Brouwer trees:
	\begin{enumerate}[(i)]
	\item $\LPO$ \label{item:snd-thm-brw-lpo} 
	\item trichotomy: $\forall x,y. (x < y) \dissum (x = y) \dissum (y < x)$ \label{item:snd-thm-trich}
	\item splitting: $\forall x,y. (x \leq y) \to (x < y) \dissum (x = y)$. \label{item:snd-thm-splitting}
\end{enumerate}
\end{theorem}
\begin{proof}
	We show $\eqref{item:snd-thm-brw-lpo}\Rightarrow\eqref{item:snd-thm-trich}\Rightarrow\eqref{item:snd-thm-splitting}\Rightarrow\eqref{item:snd-thm-brw-lpo}$.
	
	$\eqref{item:snd-thm-brw-lpo}\Rightarrow\eqref{item:snd-thm-trich}$:
	Assume $\LPO$. By \cref{thm:brw-LPO}, we can decide $x<y$.
	If it holds, we are done. Otherwise, by the contrapositive of \cref{lem:lpo-to-not-leq-lt}, we get $\neg\neg(y \leq x)$ and, since decidability implies stability, therefore $y \leq x$.
	In the same way, we decide $y < x$, which gives us either $y < x$ or $x \leq y$. The second case implies $x = y$ by antisymmetry.

	$\eqref{item:snd-thm-trich}\Rightarrow\eqref{item:snd-thm-splitting}$:
	This is an instance of \cref{lem:trich-splits}.

	$\eqref{item:snd-thm-splitting}\Rightarrow\eqref{item:snd-thm-brw-lpo}$:
	Given a binary sequence $s$, we know $\blimit \, \jump s \leq \omega \cdot 2$ from \cref{lem:jumping-omega-vs-true}.
	Splitting this inequality allows us to decide whether $\blimit \, \jump s = \omega \cdot 2$ which, again by \cref{lem:jumping-omega-vs-true}, yields the conclusion of $\LPO$. 
\end{proof}

Another natural question is if we can compute suprema of not necessarily increasing sequences of Brouwer trees. An important special case is the join or maximum $x \sqcup y$ of two trees, which is the suprema of the sequence $(x, y, y, y, \ldots)$. This is easy to compute if one of the trees is at most $\omega$, as $\omega \leq \blimit\,f$ for any $f : \N \toincr{<} \brouwer$:

\begin{theorem}[\flinkFull{https://cj-xu.github.io/agda/type-theoretic-approaches-to-ordinals/index.html\#Theorem-54}]
	If $y = n$ for a finite $n$, or $y = \omega$, we can define a function $(\_ \sqcup y) : \brouwer \to \brouwer$ calculating the binary join with $y$.
\end{theorem}
\begin{proof}
  For $y = n$ finite, we can define
  \begin{equation}
    x \sqcup n =
    \begin{cases}
      x & \text{if $n = 0$ or $x = \blimit\,f$} \\
      n & \text{if $x = 0$} \\
      \bsuc\,(x' \sqcup n') & \text{if $x = \bsuc\,x'$ and $n = \bsuc\,n'$}
    \end{cases}
  \end{equation}
  and prove that this indeed is the join of $x$ and $n$. Indeed any limit is going to be larger than any finite $n$, and $0$ is always the smallest element, hence $x \sqcup 0 = 0 \sqcup x = x$. If both Brouwer trees are successors, we know that their join is a successor as well.

  For joins with $y = \omega$, note that by \cref{thm:brw-decidable}, we can decide if $x$ is finite or not. Clearly a finite $x$ is smaller than $\omega$, and $\omega$ is the smallest infinite Brouwer tree, leading to the following definition:
  \begin{equation}
    x \sqcup \omega =
    \begin{cases}
      \omega & \text{if $x$ is finite} \\
      x & \text{otherwise}
    \end{cases}
  \end{equation}
  See our Agda formalisation for the proof that this indeed is the join of $x$ and $\omega$.
\end{proof}

This is as far as we can go --- already for $y = \omega + 1$ being able to calculate $x \sqcup y$ would imply a constructive taboo:

\begin{theorem}[\flinkFull{https://cj-xu.github.io/agda/type-theoretic-approaches-to-ordinals/index.html\#Theorem-55}]
  \label{thm:brw-lpo-to-join-to-wlpo}
  If $\LPO$ holds, then $x \sqcup (\omega + 1)$ exists for every $x : \brouwer$.
  If $x \sqcup (\omega + 1)$ exists for every $x : \brouwer$, then $\WLPO$ follows.
\end{theorem}
\begin{proof}
	Assume $\LPO$ and let $x : \brouwer$ be given. By \cref{thm:brw-LPO}, we can decide $\omega < x$.
	If this is the case, it is easy to see that $x = x \sqcup (\omega + 1)$.
	If it is not the case and $x \equiv \blimit \, f$, then every $f_i$ must be finite, implying $x = \omega$, and the binary join is $\omega + 1$. If $\neg(\omega < x)$ and $x$ is a successor, then $x$ must be finite, again allowing us to see that the join is $\omega + 1$.
	
	Now, assume that $x \sqcup (\omega + 1)$ exists for any $x$.
	We show that we can decide the equality $x = \omega$ which, by \cref{thm:brw-dec-equ}, implies $\WLPO$.
	When deciding the proposition $x = \omega$, we can assume $x \equiv \blimit \, f$, as the other cases are trivial.
	Using \cref{thm:brw-class-CFI}, we can check whether $(\blimit \, f) \sqcup (\omega + 1)$ is a successor or a limit. In the first case, any $f_i$ being infinite would lead to a contradiction, thus every $f_i$ must be finite, and $\blimit \, f = \omega$ follows.
	In the second case, we observe that $\blimit \, f = \omega$ would imply that the join with $\omega + 1$ is $\omega + 1$, yielding a contradiction.
\end{proof}

\subsection{An Alternative Equivalent Definition of Brouwer Trees} \label{subsec:alt-def-of-Brouwer}

We also considered an alternative quotient inductive-inductive construction of Brouwer trees using a path constructor
\begin{equation} \label{eq:antisym-constructor}
\mathsf{antisym} : \; x \leq y \to y \leq x \to x = y
\end{equation}
following the construction of the partiality monad~\cite{alt-dan-kra:partiality}.
This constructor should be seen as a more powerful version of the $\bbisim$ constructor, since if $f \lesssim g$, then $\blimit \, f \leq \blimit \, g$.
By Theorem 7.2.2 of the HoTT book \cite[Thm~7.2.2]{hott-book}, this constructor further implies that the constructed type is a set.
Let us write $\brouwer'$ for the variation of $\brouwer$ which uses the constructor \eqref{eq:antisym-constructor} instead of $\bbisim$.

Of course, $\brouwer'$ has antisymmetry for free, but the price to pay
is that the already very involved proof of well-foundedness becomes
significantly more difficult. After proving antisymmetry for
$\brouwer$, we managed to prove $\brouwer \simeq \brouwer'$, thus
establishing well-foundedness for $\brouwer'$ --- but we did not manage
to prove this directly.

Note that the constructor $\blimit$ of
$\brouwer$ asks for strictly increasing sequences; without that condition,
extensionality fails. 
For $\brouwer'$, one can consider removing
the condition, but $\brouwer'$ is then no longer equivalent to $\brouwer$.
Most importantly, the constructors overlap and the ability to decide whether an element is zero is lost, without which we do not know how to define e.g.\ exponentiation on $\brouwer$.


\section{Transitive, Extensional and Well-Founded Orders}
\label{sec:bookords}

%

As introduced in \cref{subsec:bookords}, $\bookord$ is the type of ordinals $(X, \prec)$ where the order is well-founded, extensional, and transitive. 
As noticed by Escard\'o~\cite{escardo:agda-ordinals},
extensionality and
Lemma 3.3 of Kraus, Escard\'o, Coquand and Altenkirch~\cite[Lem~3.3]{lmcs:3217}
imply that $X$ is a set.
It further follows that also $\bookord$ is a set~\cite[Thm~10.3.10]{hott-book}.

Clearly, the identity function is a simulation, and the composition of two (bounded) simulations is a (bounded) simulation; thus, $\leq$ is reflexive and $\leq$ as well as $<$ are transitive.
Note that the simulation requirement \eqref{item:simu-property} is a proposition by Corollary 10.3.13 of the HoTT book~\cite[Cor~10.3.13]{hott-book} (i.e.\ even if formulated using $\Sigma$ rather than $\exists$), and $X \leq Y$ is a proposition~\cite[Lem~10.3.16]{hott-book}. As a consequence, $\leq$ is antisymmetric. Similarly, if a simulation is bounded, then the bound is unique, and hence also the type $X < Y$ is a proposition. 

We recall a result of the HoTT book that we will use to prove non-constructive results:
\begin{lemma}[{\cite[Thm~10.4.3]{hott-book}}] \label{nonempty-subset-ord}
Assuming $\LEM$, $(A,\prec)$ is an ordinal if and only if every nonempty subset of $A$ has a least element.
\qed
\end{lemma}

Given ordinals $A$ and $B$, one can construct a new ordinal $A \dissum B$, reusing the order on each component, and letting $\inl(a) \prec_{A \dissum B} \inr(b)$.
This construction has been formalised by Escard\'o~\cite{escardo:agda-ordinals}, and amounts to the \emph{categorical join}. This is used in the proof of the second part of the following
lemma:

\begin{lemma} \label{ord-wrong-transitivity}
  If $A < B$ and $B \leq C$ then $A < C$. However $A \leq B$ and $B < C$ implies $A < C$ if and only if excluded middle $\LEM$ holds.
\end{lemma}
\begin{proof}
  If $A \simeq B_{\slash b}$ and $g : B \leq C$ then $A \simeq
  C_{\slash g(b)}$.  Assuming $\LEM$ and $f : A \leq B$, $g : B <
  C$, there is a minimal $c : C$ not in the image of $g \circ
  f$ by Lemma~\ref{nonempty-subset-ord} and $A \simeq C_{\slash
    c}$.  Conversely, let $P$ be a proposition; it is an ordinal with the empty
  order. Consider also the unit type $\Unit$ as an ordinal with the empty order.
  We have $\Unit \leq \Unit \dissum P$ and $\Unit \dissum P < \Unit \dissum P \dissum
  \Unit$, so by assumption $\Unit < \Unit \dissum P \dissum
  \Unit$. Now observe which component of the sum the bound is from: this shows if $P$ holds or not.
\end{proof}

%

%
%
%

As noted in the HoTT book~\cite[Thm~10.3.20]{hott-book}, $\bookord$ itself
carries the structure of an extensional well-founded order, and so is
an element of $\bookord$, albeit in the next higher universe.

\begin{theorem} \label{ord-ext-wf-transitive}
  The order $<$ on $\bookord$ is well-founded, extensional, and transitive. \qed
\end{theorem}

Since $<$ is well-founded, it is also irreflexive by \cref{lem:no-infinite-sequence}.

\begin{corollary} \label{cor:bookord-satisfies-assumptions}
	The triple $(\bookord, <, \leq)$ satisfies the assumptions (A\ref{item:assumption1}), (A\ref{item:assumption2}), and (A\ref{item:assumption3}).
\end{corollary}
\begin{proof}
	Most requirements are given by \cref{ord-ext-wf-transitive}, which in particular implies that $<$ is irreflexive.
	\cref{ord-wrong-transitivity} shows (A\ref{item:assumption3}).
	The remaining properties follow directly from the definitions.
\end{proof}

There is exactly one order on the empty type $\Empty$, and this order makes $\Empty$ into a zero for $\bookord$. Similarly there is only one irreflexive order on the unit type $\Unit$, namely the one where the only element is not related to itself. The successor of $A$ is given by $A$ adjoined with $\Unit$ to the right $\osuc{A}$, thus adding one more element greater than all the given elements. As proven by de Jong and Escard\'o~\cite{tom_smalltypes}, notably $\bookord$ has suprema of arbitrary small families of ordinals, not only of $\Nat$-indexed families, or of strictly increasing sequences, such as the case for $\brouwer$.

\begin{lemma}[\flinkPart{https://cj-xu.github.io/agda/type-theoretic-approaches-to-ordinals/index.html\#Lemma-60}]
  
  \label{thm:bookord-zersuclim}
  The type $\Empty$  is zero. The strong successor of $A$ is $\osuc{A}$, and if $F : X \to \bookord$ is an $X$-indexed family of ordinals, then its supremum $\simplesup \, F$ is the quotient $(\Sigma x : X.F x)/\sim$, where $(x, y) \sim (x', y')$ if and only if $(F x)_{\slash y} \simeq (F x')_{\slash y'}$, with $[(x, y)] \prec [(x',y')]$ if $(F x)_{\slash y} < (F x')_{\slash y'}$.
\end{lemma}
\begin{proof}
  Zero is clear. The definition of a bounded simulation implies $(X < Y) \leftrightarrow (X \dissum \Unit \leq Y)$, 
  making
  \cref{thm:calc-succ-characterisation}
  applicable.
  The definition of the type $\simplesup \, F$ can be found in the HoTT book~\cite[Lem~10.3.22]{hott-book}, and the proof that it is indeed the supremum was given by de Jong and Escard\'o~\cite[Thm~5.12]{tom_smalltypes}.
  %
\end{proof}

\begin{theorem} \label{thm:ord-addition-multiplication}
  $\bookord$ has addition given by $A + B = A \dissum B$, and multiplication given by $A \cdot B = A \times B$, with the order reverse lexicographic, i.e.\ $(x, y) \prec (x' , y')$ is defined to be $y \prec_B y' \dissum (y = y' \times x \prec_A x')$.
\end{theorem}
\begin{proof}
  The key observation is that a sequence of simulations $F0 \leq F1 \leq F2 \leq \ldots$ is preserved by adding or multiplying a constant \emph{on the left}, i.e.\ we have $C \cdot F0 \leq C \cdot F1 \leq C \cdot F2$ (but note that adding a constant on the right fails, see \cref{thm:ord-everything-undecidable} below).
  This allows us to use the explicit representation of suprema from \cref{thm:bookord-zersuclim} in the limit cases.
\end{proof}

Many constructions that we have performed for $\cnf$ and $\brouwer$ are not possible for $\bookord$, at least not constructively:

\begin{theorem}\label{thm:ord-subtraction-iff-lem}
	$\bookord$ has subtraction if and only if $\LEM$ holds.
\end{theorem}
\begin{proof}
	Let $P$ be a proposition and assume that $\bookord$ has subtraction. Then, there is $Q$ such that $P \dissum Q = \Unit$. This implies $Q \leftrightarrow \neg P$, and the assumed equation becomes $P \dissum \neg P$.
	
	For the other direction, assume $\LEM$ and let $s : X \leq Y$ be given.
	Defining $X_1 \defeq \Sigma(y: Y). \neg s^{-1}(y)$ ensures $X \dissum X_1 = Y$, where $\LEM$ is required to show that the canonical function is a simulation (equivalently, to construct the inverse).
\end{proof}

\begin{theorem} \label{thm:ord-everything-undecidable}
  Each of the following statements on its own 
  implies the law of excluded middle ($\LEM$), and each of the first five statements is equivalent to $\LEM$:
  \begin{enumerate}[(i)]
    \item The successor $(\osuc{\blank})$ 
          is $\leq$-monotone. \label{item:leq-mon}
    \item The successor $(\osuc{\blank})$ is $<$-monotone. \label{item:<-mon}
    \item $<$ is trichotomous, i.e.\ $(X < Y) \dissum (X = Y) \dissum (X > Y)$. \label{item:<-tricho}
    \item $\leq$ is connex, i.e.\ $(X \leq Y) \dissum (X \geq Y)$. \label{item:leq-connex}
    \item $\bookord$ has weak classification. \label{item:ord-weak-class}
    \item $\bookord$ has classification. \label{item:ord-class}
    \item $\bookord$ satisfies classifiability induction. \label{item:ord-CFI}
  \end{enumerate}
\end{theorem}
\begin{proof}
    We first show the chain $\LEM \Rightarrow \eqref{item:leq-mon} \Rightarrow \eqref{item:<-mon} \Rightarrow \LEM$.
    
    $\LEM \Rightarrow \eqref{item:leq-mon}$:
    Let $f: A \leq B$.
    Using $\LEM$, there is a minimal  $b : B \dissum \Unit$ which is not in the image of $f$ by Lemma~\ref{nonempty-subset-ord}.
    The simulation $A \dissum \Unit \leq B \dissum \Unit$ is given by $f \dissum b$.

    $\eqref{item:leq-mon} \Rightarrow \eqref{item:<-mon}$:
    Assume we have $A < B$. By \cref{thm:bookord-zersuclim,thm:calc-succ-characterisation}, this is equivalent to $A \dissum \Unit \leq B$.
    Assuming $(\osuc{\blank})$ is $\leq$-monotone, we get $A \dissum \Bool \leq B \dissum \Unit$, and applying \cref{thm:calc-succ-characterisation} once more, this is equivalent to $A \dissum \Unit < B \dissum \Unit$.

    $\eqref{item:<-mon} \Rightarrow \LEM$:
    Assume $P$ is a proposition. We have $\Empty < \Unit \dissum P$.
    If $(\osuc{\blank})$ is $<$-monotone, then we get $\Empty \dissum \Unit < \Unit \dissum P \dissum \Unit$. Observing if the simulation $f$ sends $\mathsf{inr}(\star)$ to the $P$ summand or not, we decide $P  \dissum  \neg P$.
    
    Next, we show $\LEM \Rightarrow \eqref{item:<-tricho} \Rightarrow \eqref{item:leq-connex} \Rightarrow \LEM$, where the first implication is given by Theorem 10.4.1 of the HoTT book~\cite[Thm~10.4.1]{hott-book}.
    
    $\eqref{item:<-tricho} \Rightarrow \eqref{item:leq-connex}$: Each of the three cases of \eqref{item:<-tricho} gives us either $X \leq Y$ or $X \geq Y$ or both.
    
    $\eqref{item:leq-connex} \Rightarrow \LEM$: Given a proposition $P$, we compare $P \dissum P$ with $\Unit$.
    If $P \dissum P \leq \Unit$ then $\neg P$, while $\Unit \leq P \dissum P$ implies $P$.
    
    The next step is to show $\eqref{item:ord-CFI} \Rightarrow \eqref{item:ord-class} \Rightarrow \eqref{item:ord-weak-class} \Rightarrow \LEM$. The first implication is \cref{thm:CFI-to-classification} and the second implication is automatic since any classifiable ordinal is also weakly classifiable.
    We have $\eqref{item:ord-weak-class} \Rightarrow \LEM$ because a classifiable proposition $P$ is either $\Empty$ (thus $\neg P$) or a successor $\osuc{X}$ (thus $P$ and necessarily $X = \Empty$) or
    the supremum of $\fstproj : \bookord_{\slash X} \to \bookord$ where there exists $X_0 < P$ (this case is impossible).

	Finally, we check $\LEM \Rightarrow \eqref{item:ord-weak-class}$.
    Thus, we assume $\LEM$ and a given $X$.
    If $X$ is the supremum of $\fstproj : \bookord_{\slash X} \to \bookord$ then either $X$ is empty or $X$ is a general limit and we are done.

    Now assume that $X$ is not the supremum of $\fstproj : \bookord_{\slash X} \to \bookord$.
    Since $X$ certainly satisfies the first part of the definition~\eqref{eq:is-sup-of}
	and we have $\LEM$ (as well as \cref{nonempty-subset-ord}) at our disposal, this means that there is 
    some $Y$ with
	\begin{equation}\label{eq:lem-thm-aux}
      \forall X'. (X' < X) \to (X'\leq Y)
    \end{equation}
    together with $\neg(X \leq Y)$ which, by the previous parts of this theorem, implies $Y < X$.
	With the terminology of \cref{subsubsec:zero-succ-sup}, this means that $Y$ is a predecessor of $X$. We show that $X$ in fact is the strong successor of $Y$.
	To do so, we additionally need that, for a given $Y'$ such that $Y < Y'$, we have $X \leq Y'$. Assume not: since $\LEM \Rightarrow \eqref{item:<-tricho}$, we then have $Y' < X$, and using \eqref{eq:lem-thm-aux} therefore $Y' \leq Y$, leading to the contradiction $Y < Y' \leq Y$.
\end{proof}



Last but not least, we consider splitting:

\begin{theorem}\label{thm:ord-split-lem}
	Inequalities in $\bookord$ split ($(X \leq Y) \to (X = Y) \dissum (X < Y)$) if and only if $\LEM$ holds.
\end{theorem}
\begin{proof}
	For the ``only if'' direction, let $P$ be a proposition; we always have $P \leq \Unit$.
	If we can split this inequality, it follows that $P \dissum \neg P$.
	
	For the other direction, we assume $\LEM$ and $e : X \leq Y$.
	If $e$ is surjective, then it is an equivalence and $X = Y$ follows.
	If $e$ is not surjective, the complement of its image has a smallest element $y_0$ by Lemma~\ref{nonempty-subset-ord}. Thus, $e$ is bounded by $y_0$ and witnesses $X < Y$.
\end{proof}

\section{Interpretations Between the Notions}
\label{sec:interpretations}

In this section, we show how our three notions of ordinals can be connected via structure preserving embeddings.

\subsection{From Cantor Normal Forms to Brouwer Trees}

The arithmetic operations of $\brouwer$ allow the construction of a function $\CtoB : \cnf \to \brouwer$ in a canonical way.
We define $\CtoB : \cnf \to \brouwer$ by:
\begin{align}
  \CtoB(\tz) &\defeq \bzero \\
  \CtoB(\tom a b) &\defeq \omega^{\CtoB(a)} + \CtoB(b)
\end{align}

\begin{theorem}[\flinkFull{https://cj-xu.github.io/agda/type-theoretic-approaches-to-ordinals/index.html\#Theorem-65}]
  \label{thm:CtoB-reflects}
  The function $\CtoB$ preserves and reflects $<$ and $\leq$, i.e., $a < b \iff \CtoB(a) < \CtoB(b)$, and  $a \leq b \iff \CtoB(a) \leq \CtoB(b)$.
\end{theorem}
\begin{proof}
  We show the proof for $<$; each direction of the statement for $\leq$ is a simple consequence.

  ($\Rightarrow$)~By induction on $a<b$. The case when $\tom a b < \tom c d$ because $a < c$ uses \cref{thm:additive-principal}.

  ($\Leftarrow$)~Assume $\CtoB(a)<\CtoB(b)$. If $a \geq b$, then $\CtoB(a) \geq \CtoB(b)$ by ($\Rightarrow$), in conflict with the assumption. Hence $a<b$ by the trichotomy of $<$ on $\cnf$.
\end{proof}

We remark once again that the above proof was only possible because of the ``correct'' definition of $\brouwer$ --- it would not be the case that $\CtoB$ preserves $<$ if we had used a ``naive'' version of Brouwer trees without path constructors.
By reflecting $\leq$ and antisymmetry, we have:

\begin{corollary}[\flinkFull{https://cj-xu.github.io/agda/type-theoretic-approaches-to-ordinals/index.html\#Corollary-66}]
  \label{cor:CtoB-injective}
  The function $\CtoB$ is injective.
\end{corollary}
\begin{proof}
	$\CtoB(a) = \CtoB(b)$ implies $\CtoB(a) \leq \CtoB(b)$ and thus, by \cref{thm:CtoB-reflects}, $a \leq b$.
	Analogously, one has $b \leq a$. Antisymmetry gives $a = b$.
\end{proof}

We note that $\CtoB$ also preserves all arithmetic operations
on $\cnf$. For multiplication, this relies on $\iota(n) \cdot \omega^x = \omega^x$ for $\brouwer$ (Lemma~\ref{thm:additive-principal}) 
--- see our formalisation for details.

\begin{theorem}[\flinkFull{https://cj-xu.github.io/agda/type-theoretic-approaches-to-ordinals/index.html\#Theorem-67}]
    \label{lem:f-arith}
    $\CtoB$ commutes with addition, multiplication, and exponentiation with base $\omega$.
\end{theorem}
\begin{proof}
As an example, we show that $\CtoB$ commutes with addition, i.e., $\CtoB(a+b) = \CtoB(a) + \CtoB(b)$ for all $a,b : \cnf$. The proof is carried out by induction on $a,b$. It is trivial when either of them is $\tz$. Assume $a = \tom x u$ and $b = \tom y v$. If $x<y$, then $a+b=b$. We have also $\omega^x < \omega^y$, which implies $\omega^{\CtoB(x)} < \omega^{\CtoB(y)}$ by Theorem~\ref{thm:CtoB-reflects}. Then by Lemma~\ref{thm:additive-principal}\eqref{item:add-prin-1} we have $\omega^{\CtoB(x)} + \omega^{\CtoB(y)} = \omega^{\CtoB(y)}$. By the same argument, from the fact $u < \omega^y$ we derive $\CtoB(u) + \omega^{\CtoB(y)} = \omega^{\CtoB(y)}$. Therefore, both $\CtoB(a+b)$ and $\CtoB(a) + \CtoB(b)$ are equal to $\CtoB(b)$. If $y \leq x$, then $a+b = \tom x {u + b}$ by definition. By the induction hypothesis, we have $\CtoB(u + b) = \CtoB(u) + \CtoB(b)$. Therefore, both $\CtoB(a+b)$ and $\CtoB(a) + \CtoB(b)$ are equal to $\omega^{\CtoB(x)} + \CtoB(u) + \CtoB(b)$.
%
%
\end{proof}

Although we cannot calculate suprema in $\cnf$, we can still ask whether $\CtoB$ preserves those that exist.
We restrict the question to the case of strictly increasing sequences.
We first note that $\CtoB$ does preserve the \emph{fundamental sequences} that we constructed for each limit CNF in the proof of \cref{lem:cnf:limit}, in the sense that $\CtoB$ sends $x$, the limit of its fundamental sequence $s$, to the limit of $\CtoB \circ s$.

\begin{lemma}[\flinkFull{https://cj-xu.github.io/agda/type-theoretic-approaches-to-ordinals/index.html\#Lemma-68}]
\label{thm:CtoB-preserves-fundamental-seqs}
  Let $x : \cnf$ be a limit, and $s$ its fundamental sequence as assigned in the proof of \cref{lem:cnf:limit}. We then have $\CtoB(x) = \blimit (\CtoB \circ s)$.
\end{lemma}
\begin{proof}
  Let $x$ be given. We analyse how the fundamental sequence $s$ of $x$ is constructed and compute, using \cref{lem:f-arith} extensively. In the following, we leave the embedding of natural numbers into both $\cnf$ and $\brouwer$ implicit, for convenience.
	\begin{enumerate}[(i)]
		\item Case $x \equiv \tom {c+1} \tz$:
		By definition, we have 
		\begin{equation}
		\begin{alignedat}{4}
		&\CtoB(\tom {c+1} \tz) &\; & = &\;\; & \omega^{\CtoB(c+1)} \\
		&&&= &&\omega^{\CtoB(c) + 1} \\
		&&&= &&\omega^{\CtoB(c)} \cdot \omega \\
		&&&= &&\omega^{\CtoB(c)} \cdot \blimit(\lambda i. i) \\
		&&&= &&\blimit (\omega^{\CtoB(c)} \cdot i)
		\end{alignedat}
		\end{equation}
		On the other hand, the fundamental sequence is in this case defined as 
		$s \defeq \lambda i.(\tom {c} \tz) \cdot i)$, and therefore
		\begin{equation}
		\begin{alignedat}{4}
		&\blimit(\CtoB \circ s) &\; & = &\;\; & \blimit\left(\lambda i. \CtoB((\tom {c} \tz) \cdot i)\right) \\
		&&&= &&\blimit\left(\lambda i. \CtoB(\tom {c} \tz) \cdot i\right) \\
		&&&= &&\blimit(\lambda i. \omega^{\CtoB(c)} \cdot i)
		\end{alignedat}
		\end{equation}

		\item Case $x \equiv \tom a \tz$, where $a$ is not a successor:
		Writing $r$ for the fundamental sequence of $a$, then the fundamental sequence of $s$ is, by definition, $\lambda i. \tom {r_i} \tz$.
		Using that $\CtoB$ preserves the limit of $r$ by induction, we have
		\begin{equation}
		\begin{alignedat}{4}
		&\CtoB(\tom {a} \tz) &\; & = &\;\; & \omega^{\CtoB(a)} \\
		&&&= &&\omega^{\blimit(\CtoB \circ r)} \\
		&&&= &&\blimit(\lambda i. \omega^{\CtoB(r_i)}) \\
		&&&= &&\blimit(\lambda i. \CtoB(\omega^{r_i})) \\ 
		&&&= &&\blimit(\CtoB \circ s). 
		\end{alignedat}
		\end{equation}	

		\item If $x = \tom a b$ with $b>0$, then $b$ necessarily is a limit with fundamental sequence $r$.
		Recall that the fundamental sequence of $x$ is then given by $\lambda i. \tom{a}{r_i}$. 
		\begin{equation}
		\begin{alignedat}{4}
		&\CtoB(\tom {a} b) &\; & = &\;\; & \omega^{\CtoB \, a} + \CtoB\, b \\
		&&&= &&\omega^{\CtoB \, a} + \blimit(\CtoB \circ r)  \\
		&&&= &&\blimit(\lambda i. \omega^{\CtoB \, a} + \CtoB(r_i))  \\
		&&&= &&\blimit(\lambda i. \CtoB (\tom a {r_i}))  \\
		&&&= &&\blimit(\CtoB \circ s).
		\end{alignedat}		
		\end{equation}
		\end{enumerate}
\end{proof}

This might seem like an encouraging first step, but in fact continuity of $\CtoB$ in general turns out to be a constructive taboo.
This is because $\cnf$ and $\brouwer$ are powerful in different ways: if $\CtoB$ were to preserve limits, then we could use the decidable equality of $\cnf$ to confirm that a CNF is the limit of some sequence, then transfer the limit across to $\brouwer$ where we could use the strong property of strict inequalities below limits factoring through one of the elements of the sequence to find an explicit witness.
Using this idea, we can show that continuity of $\CtoB$ implies Markov's principle.
Conversely, Markov's principle proves that $\CtoB$ is continuous with respect to strictly increasing sequences, so we have an exact correspondence.

\begin{theorem}[\flinkFull{https://cj-xu.github.io/agda/type-theoretic-approaches-to-ordinals/index.html\#Theorem-69}]
	$\CtoB$ preserves limits of strictly increasing sequences if and only if $\MP$ holds.
\end{theorem}
\begin{proof}
	First, assume that $\CtoB$ preserves limits of strictly increasing sequences.
	We want to show $\MP$.
	Therefore, let $s$ be a binary sequence such that $\neg(\forall i. s_i = \bff)$.
	We claim that $\omega + \omega$ is the limit of the jumping sequence $\jump s : \Nat \to \cnf$ from \eqref{eq:jumping-sequence-construction-only-one-jump-abstract}.
	The first condition we need to check for $\omega + \omega$ to be the limit is $\forall n. \jump s \, n \leq \omega + \omega$; but this is easy since every $\jump s \, n$ is either finite or of the form $\omega + k$, depending on the decidable property of whether there is $i \leq n$ with $s_i = \btt$.
	The second condition requires us to check $\forall c. (\forall n. \jump s \, n \leq c) \to \omega + \omega \leq c$.
	If $c < \omega + \omega$, then each $\jump s \, i$ is below $\omega$ and thus $s_i = \bff$, which contradicts the assumption. Therefore, we have $\omega + \omega \leq c$ thanks to the trichotomy of $\cnf$ by~\cref{thm:CNF-satisfies-general-notions}.
	
	
	By assumption, we thus have $\blimit \, (\CtoB \circ \jump s) = \omega + \omega$.
	By \cref{lem:x-under-limit-means-x-under-element}, there exists $n$ such that $\CtoB (\jump s \, (n+1)) > \omega$, and by \cref{thm:CtoB-reflects} $\CtoB$ reflects this inequality, hence $\jump s \, (n+1) > \omega$ (since $\CtoB$ preserves $\omega$ by \cref{lem:f-arith}). This means $\jump s \, (n+1) = \omega + k$ for some $k$, and indeed we must have $s_{n - k} = \btt$, i.e.,\ we have proven $\exists i. s_i = \btt$, as required.

	For the other direction, we first show the following:
	
	\textbf{Claim:} Assume we have $x,y : \cnf$ and strictly increasing sequences $f, g : \N \to \cnf$ such that $\islimof{x}{f}$ and $\islimof y g$. If $x \leq y$ and $\MP$ holds, then $f$ is simulated by $g$:
	\begin{equation}
	\forall i. \exists k. f_i \leq g_k.
	\end{equation}
	To see this, let us fix $i$ and define $h : \N \to \Bool$ by
	\begin{equation}
	h_k \defeq
	\begin{cases}
	\bff & \textit{if $g_k < f_i$}\\
	\btt & \textit{if $f_i \leq g_k$.}
	\end{cases}
	\end{equation}
	If $\forall k. h_k = \bff$, then $\forall k. g_k < f_i$ and therefore $y \leq f_i < x$ in contradiction to the assumption.
	Therefore, using $\MP$, we have $\exists k. h_k = \btt$, i.e.\ $\exists k. f_i \leq g_k$. 

	We are now ready to show that $\MP$ implies that $\CtoB$ preserves limits of strictly increasing sequences. Let $f$ be such a sequence with limit $x$; we want to show $\CtoB(x) = \blimit (\CtoB \circ f)$.
	Let $s$ be the fundamental sequence of $x$ as constructed in the proof of \cref{lem:cnf:limit}.
	Using the above claim and $\MP$, we see that $s$ and $f$ are bisimilar.
	Since $\CtoB$ preserves the relations (\cref{thm:CtoB-reflects}), $\CtoB \circ f$ and $\CtoB \circ s$ are bisimilar in $\brouwer$ and therefore have equal limits. Hence using \cref{thm:CtoB-preserves-fundamental-seqs}, we have $\CtoB(x) = \blimit (\CtoB \circ s) =  \blimit (\CtoB \circ f)$, as required.
\end{proof}

Lastly, as expected, Brouwer trees define bigger ordinals than Cantor normal forms: when embedded into $\brouwer$, all Cantor normal forms are below $\varepsilon_0$, the limit of the increasing sequence $\omega$, $\omega^{\omega}$, $\omega^{\omega^{\omega}}$, \ldots

\begin{theorem}[\flinkFull{https://cj-xu.github.io/agda/type-theoretic-approaches-to-ordinals/index.html\#Theorem-70}]
  \label{thm:cnf-below-eps0}
  For all $a : \cnf$, we have $\CtoB(a) < \blimit\,(\lambda k.\omega \uparrow\uparrow k)$, where $\omega \uparrow\uparrow 0 \defeq \omega$ and $\omega \uparrow\uparrow (k+1) \defeq \omega^{\omega \uparrow\uparrow k}$.
\end{theorem}
\begin{proof}
  By induction on $a$. Using that $\varepsilon_0 = \omega^{\varepsilon_0} = \omega^{\omega^{\varepsilon_0}}$, in the step case we have $\omega^{\CtoB(a)} + \CtoB(b) < \varepsilon_0$ by \cref{thm:additive-principal}, strict monotonicity of $\omega^-$, and the induction hypothesis.
\end{proof}

\subsection{From Brouwer Trees to Extensional Well-Founded Orders}

As $\brouwer$ comes with an order that is well-founded, extensional, and transitive, it can itself be seen as an element of $\bookord$.
Every ``subtype'' of $\brouwer$ (constructed by restricting to trees smaller than a given tree) inherits this property, giving a canonical function from Brouwer trees to extensional, well-founded orders. We define
\begin{equation}
\BtoO(a) = \Sigma(y : \brouwer).(y < a).
\end{equation}
with order relation $(y, p) \prec (y', p')$ if $y < y'$. This extends to a function $\BtoO : \brouwer \to \bookord$.
The first projection gives a simulation $\BtoO(a) \leq \brouwer$:

\begin{lemma} \label{lem:projection-simulation}
	For $X : \bookord$ with $x : X$, the first projection $\fstproj : X_{\slash x} \to X$ is a simulation.
	If $x, y : X$ and $f : X_{\slash x} \to X_{\slash y}$ is a function, then $f$ is a simulation if and only if $\fstproj \circ f = \fstproj$.
\end{lemma}
\begin{proof}
	Both properties required in the definition of a simulation are obvious in the case of $\fstproj$.
	In the second sentence, if $f$ is a simulation, then the equality follows from the uniqueness of simulations~\cite[Thm~10.3.16]{hott-book}.
	If the equality holds then, again, the two properties in the definition of a simulation are clear for $f$.
\end{proof}

Using extensionality of $\brouwer$, this implies that $\BtoO$ is an embedding from $\brouwer$ into $\bookord$. Using that $<$ on $\brouwer$ is propositional, and that
carriers of orders
are sets, it is also not hard to see that $\BtoO$ is
order-preserving:

\begin{lemma}[\flinkPart{https://cj-xu.github.io/agda/type-theoretic-approaches-to-ordinals/index.html\#Lemma-72}]
  \label{lem:BtoO-injective}
  The function $\BtoO : \brouwer \to \bookord$ is injective, and preserves  $<$ and $\leq$.
\end{lemma}
\begin{proof}
	The first part (injectivity of $\BtoO$) is a special case of the following statement: Given $X : \bookord$, the map $X \to \bookord$, $x \mapsto X_{\slash x}$ is injective. 
	This is remarked just before Definition 10.3.19 in the HoTT book~\cite[Def~10.3.19]{hott-book}. We give a detailed proof:
	
	Note that an equality $Y = Z$ in $\bookord$ gives rise to a canonical simulation $X \leq Y$ by path induction.
	Now, assume $x,y : X$ with $X_{\slash x} = X_{\slash y}$.
	We get $f : X_{\slash x} \leq X_{\slash y}$.
	By \cref{lem:projection-simulation}, $f$ maps $(z,p)$ to $(z,q)$, with $q : z < y$; that is, every element below $x$ is also below $y$.
	The symmetric statement follows by the symmetric argument, and injectivity of $x \mapsto X_{\slash x}$ by extensionality.
	
	If $q : x \leq y$ in $\brouwer$, then the map $X_{\slash x} \to X_{\slash y}$, $(z,p) \to (z, p \cdot q)$ is a simulation by \cref{lem:projection-simulation}, thus $\BtoO$ preserves $\leq$.
	This implies that $<$ is preserved as well since $X < Y \leftrightarrow (X \dissum \mathsf 1) \leq Y$.
\end{proof}

A natural question is whether the above result can be strengthened further, i.e.\ whether $\BtoO$ is a simulation.
David W\"arn pointed out to us that this is a special case of the map $x \mapsto X_{\slash x} : X \to \bookord$ being a simulation for any small ordinal $X$.

\begingroup
\makeatletter
\apptocmd{\thetheorem}{\unless\ifx\protect\@unexpandable@protect\protect\footnote{\cref{thm:lem-implies-simulation,rem:published-paper-had-wrong-statement} are updated compared to the \href{https://doi.org/10.1016/j.tcs.2023.113843}{publisher's version}.}\fi}{}{}
\makeatother

\begin{theorem}\label{thm:lem-implies-simulation}
  The function $\BtoO : \brouwer \to \bookord$ is a simulation.
\end{theorem}
\begin{proof}
  Given $B < \BtoO(a)$, we need to find a Brouwer tree $a' < a$ such that $\BtoO(a') = B$. By definition of $B < \BtoO(a)$, there is $(a', p) :  \Sigma(y : \brouwer).(y < a)$ such that $B \simeq \BtoO(a)_{\slash (a', p)}$. Since $a' < a$, we have $\BtoO(a)_{\slash (a', p)} = {\Sigma(y : \brouwer)}.((y < a) \times (y < a')) \simeq \BtoO(a')$, and hence $\BtoO(a') = B$ as needed.
\end{proof}
\endgroup

\begingroup
\addtocounter{footnote}{-1}\addtocounter{Hfootnote}{-1}
\makeatletter
\apptocmd{\thetheorem}{\unless\ifx\protect\@unexpandable@protect\protect\footnotemark\fi}{}{}
\makeatother

\begin{remark}\label{rem:published-paper-had-wrong-statement}
  Unfortunately, in an earlier version of this paper, we erroneously claimed that $\BtoO : \brouwer \to \bookord$ being a simulation would imply the constructive taboo of $\WLPO$. We are thankful to David W\"arn for catching our mistake.
\end{remark}

\endgroup

We trivially have $\BtoO(\bzero) = \Empty$. 
One can further prove that $\BtoO$ commutes with limits, i.e.\ 
$\BtoO(\blimit(f)) = \osup{(\BtoO \circ f)}$.
However, $\BtoO$ does \emph{not} commute with successors;
it is easy to see that $\BtoO (x) \dissum \Unit \leq \BtoO(\bsuc \, x)$, but the other direction 
implies $\WLPO$.
This also means that $\BtoO$ does not preserve the arithmetic operations but ``over-approximates''
them, i.e.\ $\BtoO(x + y) \geq \BtoO (x) \dissum \BtoO (y)$
and
$\BtoO(x \cdot y) \geq \BtoO (x) \times \BtoO (y)$.

\section{Computational Efficiency of Our Notions of Ordinals}
\label{sec:bench}

\begin{figure}
  \centering
  \includegraphics[width=\textwidth]{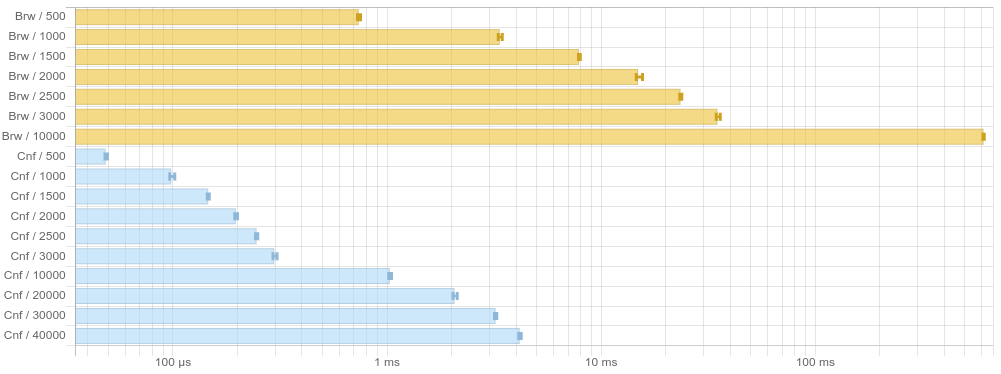}
  \caption{Benchmarking running time for $H_{\omega^n}(1)$, $n = 500$, $1000$, \ldots, for Brouwer trees (Brw) and Cantor normal forms (Cnf).}
  \label{fig:hardy}
\end{figure}

Apart from logical expressiveness, it is also interesting to compare
the computational efficiency of our different notions of ordinal. This
is possible to do, since we have formalised them in Cubical Agda,
which has computational support for higher inductive types and the
Univalence Axiom. Inspired by Berger's benchmarking of ordinal
recursive versus higher type programs extracted from Gentzen's proof
of transfinite induction up to $\varepsilon_0$~\cite{Berger01a}, we
compared the efficiency of our different ordinal representations for
computing $H_{\omega^n}(1)$, where, for each notion of ordinal $\O$, $H : \O \to \N \to \N$ is
the Hardy hierarchy~\cite{hardy}, with
\begin{align}
  &H_0(n) = n \\
  &H_{\alpha+1}(n) = H_{\alpha}(n+1)\\
  &H_{\lim{f}}(n) = H_{f(n)}(n).
\end{align}
However, since obviously $H_{\lim{f}}$ depends on the choice of
fundamental sequence $f$ in the limit case, this is not a well defined
function on ordinals. To work around this issue, we instead compute
 $H : \O \to \N \to \proptrunc{\N}$, where $\proptrunc{\N}$ is the propositional truncation of $\Nat$. All elements of $\proptrunc{\N}$ are propositionally equal, but we can still let Cubical Agda compute their normal form, which will be of the form $\truncP{k}$ for some numeral $k$, which we can extract for a closed program. We are thus interested in the following defining equations:
\begin{align}
  &H_0(n) = \truncP{n}\\
  &H_{\alpha+1}(n) = H_{\alpha}(n+1)\\
  &H_{\lim{f}}(n) = H_{f(n)}(n)
\end{align}
For $\O = \brouwer$, this definition can now be implemented directly
by induction on the Brouwer tree, whereas for $\O = \cnf$, we use
classifiability induction to define $H$. We cannot define $H$ at all
for $\O = \bookord$, since classification for $\bookord$ is a
constructive taboo. Using Cubical Agda's \texttt{-{}-erased-cubical}
feature, we compiled these definitions and ran $H_{\omega^n}(1)$
for increasing values of $n$ --- the result is the same for each $n$,
but the run time increases. The results can be found in Figure~\ref{fig:hardy}.  As can be seen there, Cantor normal forms are
significantly more efficient than Brouwer trees for this
computation. This could be in part due to their first-order
representation, but also perhaps due to our implementation of
classifiability induction for Cantor normal forms: this follows
Gentzen's proof of transfinite induction up to $\varepsilon_0$, and as
Berger~\cite{Berger01a} noticed, this gives rise to an
efficient, higher-order implementation.

\section{Conclusions and Future Directions}

We have introduced Cantor normal forms ($\cnf$), Brouwer trees ($\brouwer$), and extensional well-founded orders ($\bookord$), three different approaches to ordinal theory in the setting of homotopy type theory. Even though these approaches are quite different in their implementation, we have shown that they can all be studied in a single abstract setting and shown to share many expected properties of ordinals from their classical theory.
It is our hope that our work may shed light on other constructive or formalised approaches to ordinals also in other settings~\cite{BFT:nested:mset,BPT:card:isa,MV:ord:acl2,schmitt:ord:key}.

Cantor normal forms are a formulation where most properties are decidable, while the opposite is the case for extensional well-founded orders.
Brouwer trees sit in the middle, with some of its properties being decidable, such as being a finite ordinal. However other properties, such as deciding equality between Brouwer trees in general, is a constructive taboo in the sense that it is equivalent to the non-constructive principle $\LPO$. This is in contrast to the situation for extensional well-founded orders, where decidability of most properties is equivalent to the constructively much stronger principle $\LEM$. In the future, we plan to investigate such decidability aspects further, including the notion of \emph{semidecidability}~\cite{tom:semidec_in_Agda} from synthetic computability theory~\cite{bauer2006first,forster2019synthetic}. For example, if $c : \cnf$ is smaller than $\omega^2$, then the
families $(\CtoB \, c \leq \blank)$ and $(\CtoB \, c < \blank)$ are
semidecidable.

Along another dimension, the canonical maps $\CtoB : \cnf \to \brouwer$ and $\BtoO : \brouwer \to \bookord$ embeds ``smaller'' types of ordinals into ``larger'' ones: while every element of $\cnf$ represents an ordinal below $\varepsilon_0$, $\brouwer$ can go much further, and since $\brouwer$ can be viewed as an element of $\bookord$, the latter can clearly reach larger ordinals than the former by the Burali-Forti argument~\cite{escardo_et_al:BuraliForti,burali1897questione}. To at least partially overcome these limitations comparing $\cnf$ to $\brouwer$, it would be interesting to consider more powerful ordinal notation systems such as those based on the Veblen functions~\cite{schuette, veblen} or collapsing functions~\cite{bachmann,buchholz}, and see how these compare to Brouwer trees.

One can also explore more powerful variations of Brouwer trees. Following Schwichtenberg's approach~\cite{schwichtenberg:leeds90}, we could replace limits of countable sequences with larger limits and construct higher number classes as quotient inductive-inductive types in a similar way, e.g.\ a type $\brouwer_2$ closed under limits of $\brouwer$-indexed sequences, and then more generally types $\brouwer_{n+1}$ closed under limits of $\brouwer_n$-indexed sequences, and so on.


Finally, there are interesting connections between the ordinals we can
represent and the proof-theoretic strength of the ambient type theory:
each proof of well-foundedness for a system of ordinals is also a lower
bound for the strength of the type theory it is constructed in. It is
well known that definitional principles such as simultaneous
inductive-recursive definitions~\cite{IRstrength} and higher inductive
types~\cite{lumsdaine:hits} can increase the proof-theoretical
strength, and so, we hope that they can also be used to faithfully
represent even larger ordinals.

\section*{\emph{Acknowledgements}}

We would like to thank
the participants of the conferences MFCS'21,
TYPES'21, TYPES'22, CCC'22, MGS'22, and PC'22 for their comments and suggestions related to this work.
In particular, we are grateful to
Tom de Jong, Helmut Schwichtenberg, Thorsten Altenkirch, Mart\'in H\"otzel Escard\'o, Peter Hancock, Paul Taylor, Andreas Abel, and David W\"arn
for insightful discussions.
We also thank the anonymous reviewers of the current paper and its previous MFCS'21 version for their comments.

Funding: This work was supported by The Royal Society (grant reference URF\textbackslash{}R1\textbackslash{}191055) and the UK National Physical Laboratory Measurement Fellowship project ``Dependent types for trustworthy tools''.


\bibliographystyle{plain}
\bibliography{references}

\end{document}